\documentclass[dvipsnames,twocolumn]{article}
\let\oldput\put
\def\put(#1,#2)#3{%
  \oldput(#1,#2){\sffamily #3}%
}

\usepackage[letterpaper, total={190mm,240mm},top=20mm]{geometry}
\usepackage{tikz}
\usepackage{url} %this package should fix any errors with URLs in refs.
\usepackage{enumitem}
\usepackage{amsmath}
\usepackage{overpic}
\usepackage{pict2e}
\usepackage{picture}
\usepackage[switch]{lineno}
\usepackage{authblk}
\usepackage{amssymb}
\usepackage{overpic}
\usepackage{nicefrac}
\usepackage{lipsum, babel}
\usepackage[super,numbers,sort&compress]{natbib}

\title{Mid-troposphere moisture driving the onset of convective self-aggregation}
\title{Convective self-aggregation through cold pool interaction}
\title{Convective self-aggregation facilitated by high resolution}
\title{High resolution convective self-aggregation facilitated by the diurnal cycle}
\title{Can the diurnal cycle resolve the convective self-aggregation resolution dilemma?}
\title{The diurnal cycle can resolve the convective self-aggregation resolution dilemma}
\title{Solving the convective self-aggregation resolution dilemma with a diurnal cycle}
\title{Shaking up the convective self-aggregation paradox}
\title{Unexpected convective aggregation at high spatial resolution}
\title{Persistent tropical aggregation from mesoscale convective systems}
\title{Emergent self-aggregation from mesoscale convective systems}
\title{Tropical thunderstorm aggregation induced by strong cold pools}
\title{Convective self-aggregation triggered by strong cold pools}
\title{Convective self-aggregation triggered by mesoscale convective systems}
\title{Convective self-aggregation evoked by cloud interactions} 
\title{Convective self-aggregation triggered by diurnal mesoscale convective systems}
\title{On the diurnally-enhanced cloud clustering and mesh resolution?}
\title{The diurnal path to convective self-aggregation}
\title{The diurnal route to persistent convective self-aggregation}
\title{The diurnal path to persistent convective self-aggregation}

\author[1]{Gorm Gruner Jensen}
\author[1]{Romain Fi\'evet}
\author[1,2,3]{Jan O. Haerter}
\affil[1]{Niels Bohr Institute, Copenhagen University, Blegdamsvej 17, 2100 Copenhagen, Denmark}
\affil[2]{Complexity and Climate, Leibniz Center for Tropical Marine Research, Fahrenheitstrasse 6, 28359 Bremen, Germany}
\affil[3]{Jacobs University Bremen, Campus Ring 1, 28759 Bremen, Germany}

\begin{document}

\maketitle

%\begin{abstract}
% ****************
% *** ABSTRACT ***
% ****************
%\noindent
\noindent
{\bf
Clustering of tropical thunderstorms constitutes an important climate feedback because it influences the heat radiated to space.
Convective self-aggregation (CSA) is a profound modelling paradigm for explaining the clustering of tropical oceanic thunderstorms.
However, CSA is hampered in the realistic limit of fine model resolution when cold pools---dense air masses beneath thunderstorm clouds---are well-resolved. 
Studies on CSA usually assume the surface temperature to be constant, despite realistic surface temperatures varying significantly between night and day, even over the sea.
Here we mimic oscillating surface temperatures in cloud resolving numerical experiments and show that, in the presence of a diurnal cycle, CSA is enabled by high resolutions.
We attribute this finding to vigorous combined cold pools emerging in symbiosis with mesoscale convective systems. 
Such cold pools suppress buoyancy in extended regions ($\mathbf{\sim 100}$ km) and enable the formation of persistent dry patches.
Our findings help clarify how the tropical cloud field forms sustained clusters under realistic conditions and may have implications for the origin of extreme thunderstorm rainfall and tropical cyclones.
}

% ********************
% *** Introduction ***
% ********************
\rmfamily % Use serif font for main text

CSA refers to the spatial separation into deep convective and dry subregions occurring spontaneously in numerical simulations with homogeneous boundary and initial conditions
%"spontaneous spatial organization of convection in numerical simulations of radiative convective equilibrium despite homogeneous boundary conditions and forcing,"
\cite{held1993radiative, tompkins1998radiative, bretherton2005energy,wing2017convective}. 
CSA serves as a plausible mechanism for observed large-scale tropical convective clustering, including the Madden-Julian oscillation\cite{zhang2005madden} or the formation of tropical cyclones.\cite{emanuel2018100}
Modelling suggests that CSA typically hinges on local radiation feedbacks\cite{muller2012detailed,emanuel2014radiative,muller2015favors,coppin2015physical,hohenegger2016coupled}.
%The final state of CSA is characterised by a clear system-scale separation between a dry, rain-free subsidence region and a moist area with intense deep convective (thunderstorm) activity.
Maintenance of %the system-scale separation seen in 
CSA has been attributed to a large-scale circulation resulting in an upgradient moisture transport\cite{craig2013coarsening,emanuel2014radiative,muller2015favors,holloway2016sensitivity}.
The circulation is driven by a combination of moist adiabatic lifting in the convectively active region and enhanced radiative cooling in the dry region which must be compensated by subsidence heating.
The initial harbinger of CSA is the formation of several small persistent dry patches \cite{wing2017convective}.
At this initial stage, low cloud \cite{muller2012detailed} and moisture feedbacks \cite{emanuel2014radiative,muller2015favors} within dry regions were found to be critical for overcoming the re-distribution of moisture by negative feedbacks \cite{bretherton2005energy}.
Cold pools (CPs)---density currents produced by rain re-evaporation beneath thunderstorm clouds---were reported to act against such clustering\cite{jeevanjee2013convective,muller2015favors,boye2019self,yanase2020new}.
Also finer horizontal grid resolution, which intensifies CP effects \cite{muller2012detailed,moseley2020impact, hirt2020cold}, hampered the onset.

Whereas CSA studies often specialise to temporally constant boundary conditions\cite{wing2017convective}, it is observationally evident that also temporal variations influence the spatial characteristics of convective rainfall \cite{chen1997diurnal,dai2001global,kawai2007diurnal,suzuki2009diurnal,bellenger2009analysis,bellenger2010role, peatman2014propagation}. 
% Indeed, over continents, where surface temperatures oscillate strongly over the day, a large fraction of extreme rainfall results from mesoscale convective systems (MCS) \cite{tan2015increases, schumacher2020formation}.
Indeed, over continents, where surface temperatures oscillate strongly between day and night, a large fraction of extreme rainfall results from mesoscale convective systems (MCSs) \cite{tan2015increases, schumacher2020formation}. 
MCS are defined as thunderstorm clusters exceeding $100$ kilometres spatially and three hours temporally\cite{houze2004mesoscale}.
%Despite the fact that much of tropical rainfall extremes stem from MCS\cite{tan2015increases} and 
Despite indications that MCS rainfall rates and volumes might be increasing \cite{westra2014future,prein2017simulating}, the forecast performance for MCS remains low \cite{fritsch2004improving,sukovich2014extreme}.

%To mimic diurnal variation, simulation efforts have %aimed at understanding how convective self-aggregation is influenced, when the 
%relaxed the requirement of constant boundary conditions
Several studies mimic diurnal variation through oscillating surface temperatures
\cite{liu1998numerical,tian2006modulation,cronin2015island,ruppert2018diurnal,ruppert2019diurnal}.
Under such conditions, recent simulations demonstrated spontaneous formation of MCS-like clusters %in repeated diurnal cycle simulations 
\cite{haerter2020diurnal},
%spatial patterns, which consist of extended convective clusters spontaneously forming within a single day and persisting over several hours \cite{haerter2020diurnal}.
% These clusters
% %, which were classified as mesoscale convective systems\cite{houze2004mesoscale} (MCS), 
% were found to appear only when the amplitude of the diurnal cycle was sufficiently large ($\gtrsim 3.5\,K$) and were attributed to vigorous "combined cold pools," which were able to force moist boundary layer air to the level of free convection.
% When the amplitude was smaller ($\approx 2\,K$), neither MCSs nor combined cold pools were detected and the organisational pattern was similar to the near-random pattern during early stages of radiative-convective equilibrium (RCE) simulations.
% The former, large-surface temperature amplitude organisational pattern was referred to as "diurnal self-aggregation" (DSA).
% DSA was described as similar to convective self-aggregation in that deep convective clustering spontaneously develops and is concentrated in parts of the spatial domain, but DSA differs from CSA in a number of aspects:
% most importantly, %the diurnal clusters did not persist from one day to the next.
% the clusters organise into patterns that are anti-correlated from day to day, where an area receiving pronounced rain on a given day will typically %be dry on the subsequent day.
which appeared only when the surface temperature amplitude was sufficiently large ($\gtrsim 3.5\,K$). 
The clusters were attributed to vigorous "combined cold pools," which were able to force moist boundary layer air to the level of free convection.
When the amplitude was smaller ($\lesssim 2\,K$), neither MCSs nor combined cold pools were detected and the organisational pattern was similar to the near-random pattern during early stages of radiative-convective equilibrium (RCE) simulations.
The organisational pattern observed in the presence of a large diurnal amplitude---referred to as "diurnal self-aggregation" (DSA)---is similar to CSA in that clustering occurs spontaneously and is concentrated in parts of the spatial domain.
However, DSA differs from CSA as clusters organise into patterns that are anti-correlated from day to day, 
such that an area receiving pronounced rain on a given day will typically receive weak rain, or none at all, on the following.

Here, we use a simulation setup that yields classical CSA under constant boundary conditions, when resolution is coarse, but fails to yield CSA when resolution is fine.
We find that for time-varying boundary conditions the alternating day-to-day dynamics is accompanied by a persistent pattern of dry and rain-free patches. 
The locked-in patches closely resemble those at the onset of classical CSA. %(\emph{i.e.} with a constant surface temperature).
%The dry anomalies span the entire depth of the troposphere and show significantly increased radiative cooling.
%Further, when the diurnal cycle is removed, these dry patches persist and deepen, despite the simulations running at parameters that do not permit CSA in the pure RCE case.
Contrary to CSA, these dry patches preferably occur at high numerical resolutions.
Thus, our results introduce an organisational mechanism that is relevant in the realistic limit of high spatial model resolution and we draw a connection between the origin of continental extreme events, caused by MCS, and that of persistent cloud clumping over the tropical ocean.

%\section{Results}
%\subsection{Spatial organisation of rainfall}
\begin{figure*}
    \centering
    \begin{overpic}{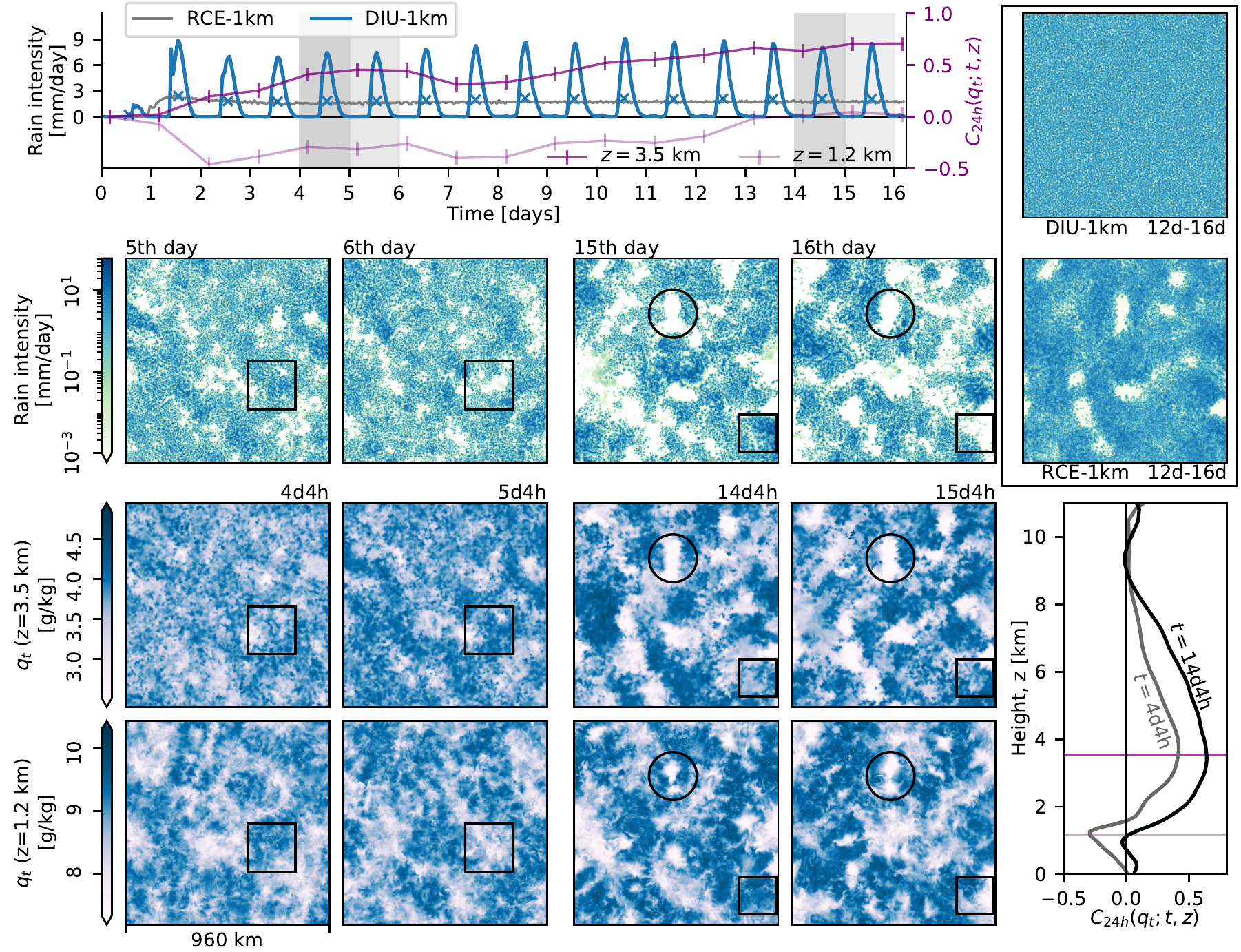}
    \put(1,76){\bf a}
    %\put(30,53){\bf b}
    %\put(46,53){\bf c}
    %\put(62,53){\bf d}
    \put(1,56){\bf b}
    \put(1,36){\bf c}
    \put(1,17){\bf d}
    
    \put(81,74){\bf e}
    \put(81,54){\bf f}
    \put(81,35){\bf g}
    
    % Add horizontal scale arrows 
    \linethickness{1pt}	
    \put(10,1){\color{black}\line(1,0){5}}   % Left
    \put(10,1){\color{black}\line(1,0.3){1}}
    \put(10,1){\color{black}\line(1,-0.3){1}}
    \put(26.5,1){\color{black}\line(-1,0){5}}   % Right
    \put(26.5,1){\color{black}\line(-1,0.3){1}}
    \put(26.5,1){\color{black}\line(-1,-0.3){1}}
    % adding one more arrow on top of all
    %  \linethickness{1.4pt}	
    % \put(-15,128){\color{black}\line(1,0){30}}
    % \put(15,128){\color{black}\line(-1,0.4){2}}
    % \put(15,128){\color{black}\line(-1,-0.4){2}}
    % \put(-22,127){DIU}
    % \put(17,127){RCE}    
    
    \end{overpic}
    \caption{
    \textbf{Spatio-temporal organisation by diurnal surface temperature oscillations.}
    \textbf{a}, Time series of domain-mean rain intensity in RCE-1km (grey curve) and DIU-1km (blue curve). 
    Blue $\times$-symbols indicate daily-average rain intensity in DIU-1km.
    %The grey curve show the rain intensity of a constant RCE-1km for comparison.
    Dark and faint purple curves show time series of 24-hour Pearson correlations, $C_{24h}(q_t;t,z)$ for total water mixing ratio, $q_t$,
    for $z=3.5$~km and $z=1.2$~km respectively, with $t$ taken at 4h on any given day ({\it Details:} Methods).
    \textbf{b}, Daily surface rainfall intensity, temporally-averaged over the 5th, 6th, 15th and 16th day. %days as labelled. 
    Corresponding averaging periods indicated by grey shades in (a).
    \textbf{c}, Early-morning (4h) horizontal field of $q_t(t,x,y,z)$ for $z=3.5$~km at times corresponding to the days in (b). % for various simulation days, as labelled. 
    \textbf{d}, Analogous to (c) but for $z=1.2$~km.
    Black squares and circles in (b)---(d) highlight regions discussed in the main text.
    \textbf{e}, Four-day average rain intensity from day 13---16 in RCE-1km. 
    The colour scale is the same as for panel (b). 
    Note the lack of spatial organisation.
    \textbf{f}, Analogous to (e), but for DIU-1km. 
    Note the rain-free patches.
    \textbf{g}, %$C(q_t(z,t),q_t(z,t+24h))$ 
    $C_{24h}(q_t;t,z)$ vertical profiles, with $t=$4d4h (grey) and $t=$14d4h (black), respectively. 
    The horizontal dark and faint purple lines indicate the respective vertical levels used in (a), (c) and (d).
    \label{fig:DIU-1km}
    }
\end{figure*}
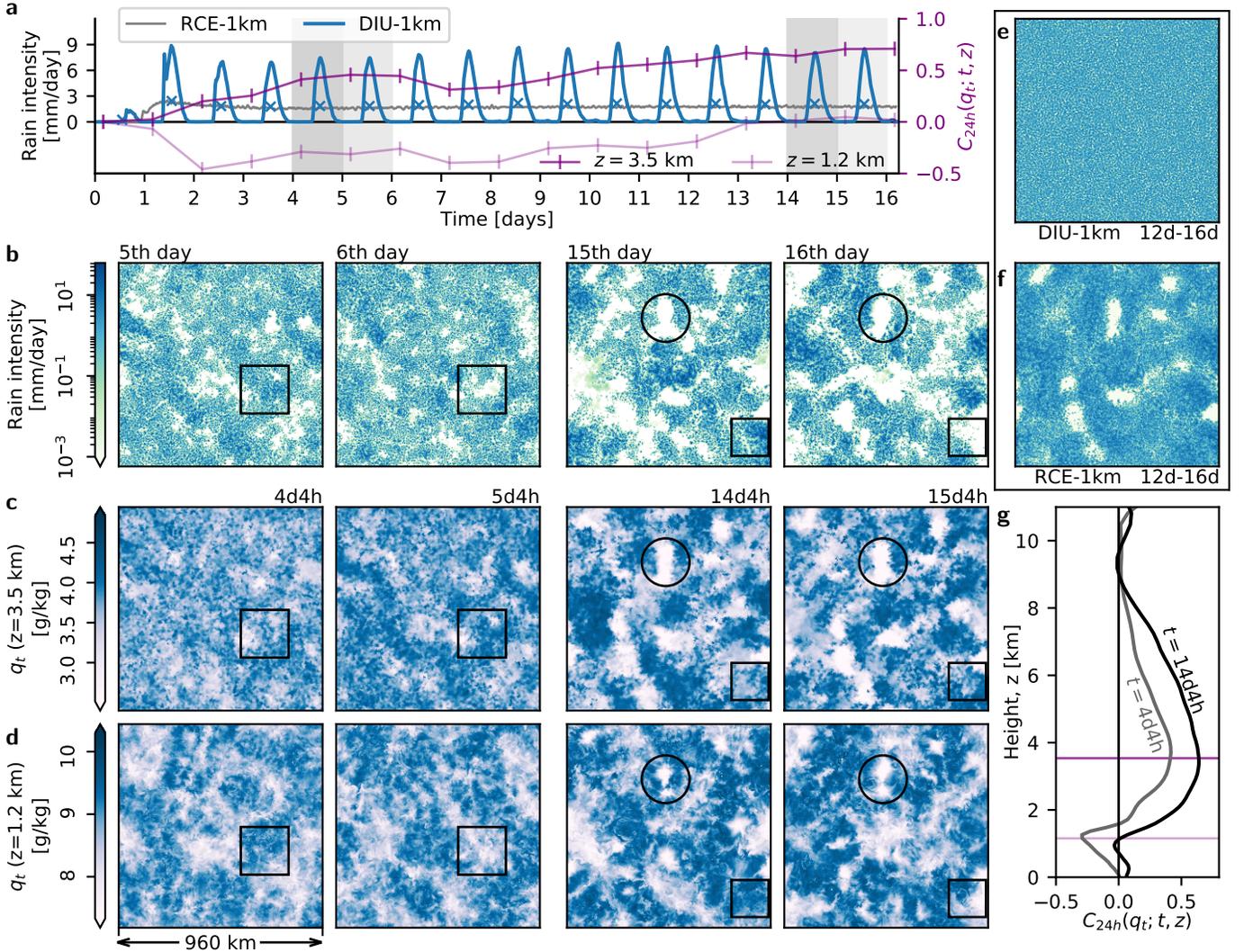

\noindent
{\it Numerical experiments:}
We study cloud resolving numerical experiments using horizontally square domains of linear size $L$ with laterally periodic boundary conditions at horizontal resolutions $dx$ of .5, 1, 2 and 4 kilometres.
%The respective domain sizes 
Domain size is chosen as $L=480$~km for $dx=.5$~km and $L=960$~km otherwise (Tab.~\ref{tab:experiments}).
In our diurnal cycle experiments --- termed DIU --- we prescribe a spatially uniform, but temporally harmonic surface temperature, $T_s(t)$, defined by a one day period and a $\pm 5$ K amplitude, such that $T_s(t)$ oscillates around the average temperature $\overline{T}_s\equiv 298$ K.
To mimic a forested land surface, we reduce surface latent heat fluxes to 70 percent of the value for a sea surface.
%$$  T_s = \overline{T}_s - \delta T_s\cos\left(\frac{2\pi t}{24h}\right) $$
Insolation peaks at noon and vanishes at night, but plays a minor role due to the prescribed surface temperature.
%and a half-sinusoidal insolation:
%$$  S = \max\Big(0, - \frac{\overline{S}}{\pi}\cos\left(\frac{2\pi t}{24h}\right)\Big) $$
Each experiment is paired by a control experiment --- termed RCE --- where all settings are preserved except that surface temperature and insolation are both kept constant at % their respective time average ({\it Details}: Methods).
the respective time averages ({\it Details}: Methods).

% \noindent
% {\bf Spontaneous dry patch formation.}
% \subsection*{Spontaneous dry patch formation}
\subsection*{Two layers of convective organization}
Consider DIU and RCE at 1~km resolution, termed DIU-1km and RCE-1km, respectively (Fig.~\ref{fig:DIU-1km}).
After a short spin-up period, rain intensity remains nearly constant in RCE-1km.
In DIU-1km, the oscillating $T_s(t)$ is reflected in oscillations in domain mean rain intensity, with a pronounced afternoon peak and mostly rain-free nocturnal conditions (Fig.~\ref{fig:DIU-1km}a). 
Daily-mean rainfall is, however, very similar in the two experiments.

% As found previously\cite{haerter2020diurnal}, during the first few days the rainfall develops a patchy mesoscale pattern with rain clusters measuring on the order of one hundred kilometres across (Fig.~\ref{fig:DIU-1km}b, 5th and 6th day).
% RCE-1km shows no signs of CSA, and the rainfall pattern is almost perfectly homogeneous
% at length scales above a few kilometers---the typical size of individual deep convective events 
% The spatial pattern emerging in DIU-1km is strikingly different:
Whereas RCE-1km shows no sign of CSA (Fig~\ref{fig:DIU-1km}e),
the spatial pattern emerging in DIU-1km is strikingly different:
during the first few days the rainfall develops a patchy mesoscale pattern with rain clusters measuring on the order of one hundred kilometres across (Fig.~\ref{fig:DIU-1km}b, 5th and 6th day), in compliance with previous findings\cite{haerter2020diurnal}.
However, ten days later (days 15 and 16), the dynamics follows a more complex pattern:
in addition to the now larger and more intense MCSs the domain is also spotted with persistently rain-free patches.
This persistent lack of rainfall becomes particularly clear, when considering rainfall averaged during multiple days (Fig~\ref{fig:DIU-1km}f).

To generate the patterning found in DIU-1km, two separate mechanisms appear to be active: 
(i) a negative feedback inhibiting convective activity in areas where rain was particularly abundant the day before;
(ii) a positive feedback which can preserve inactivity from day to day.
To analyse these feedbacks further, we turn to the horizontal moisture field $q_t(z,t)$ at each vertical model level $z$ and describe its temporal evolution from day to day using the 24h-lag correlation $C_{24h}(q_t;t,z)$ ({\it Details:} Methods).
Early morning is chosen as a reference, because at this time the moisture field is diffusively smoothed due to the absence of convective activity.
Examining the vertical profile of $C_{24h}(q_t;t,z)$ reveals an interesting dynamical structure of two pronounced extrema (Fig.~\ref{fig:DIU-1km}g): 
a global minimum near 1.2~km and a global maximum near 3.5~km, indicative of alternating moisture patterns at the lower, but persistent moisture patterns at the upper level.
In the course of the simulation, the correlations generally increase towards more positive values and by day 14 even the minimum at 1.2~km appears to switch to positive values.

The moisture pattern at this lower level, which corresponds approximately to cloud base, closely mirrors that of the rainfall ({\it compare} Fig. \ref{fig:DIU-1km}b and d).
Examples are highlighted by black squares:
at 4d4h, there is a strong positive moisture anomaly at z=1.2~km.
On the following (fifth) day that region receives intense rainfall, resulting in an intense drying near the cloud base (at 5d5h).
On the sixth day, the area receives almost no rainfall.

The positive 24h-lag correlation in the free troposphere ($z=3.5$~km) implies persistence from day to day. 
Indeed, inspecting an example of a persistently dry patch (circled in Fig. \ref{fig:DIU-1km}b,c,d), rainfall is absent in the same region during consecutive days.
Rather than replenishing the moisture within the persistently dry patches, rain clusters now appear to transport moisture elsewhere, undergoing a day-to-day oscillatory dynamics that specifically avoids the dry patches.
Hence, despite the initial day-to-day alternation in rainfall pattern, later days show sustained rain-free sub-regions --- suggesting that a form of CSA might eventually emerge in DIU-1km.
% The features described become even more pronounced for higher resolution (Fig.~\ref{fig:DIU-500m} vs. Fig.~\ref{fig:DIU-1km}).
The persistent dry patches described here emerge even faster for in DIU-500m (Fig.~\ref{fig:DIU-500m} vs. Fig.~\ref{fig:DIU-1km}).

\begin{figure*}[ht]
    \centering
    \begin{overpic}{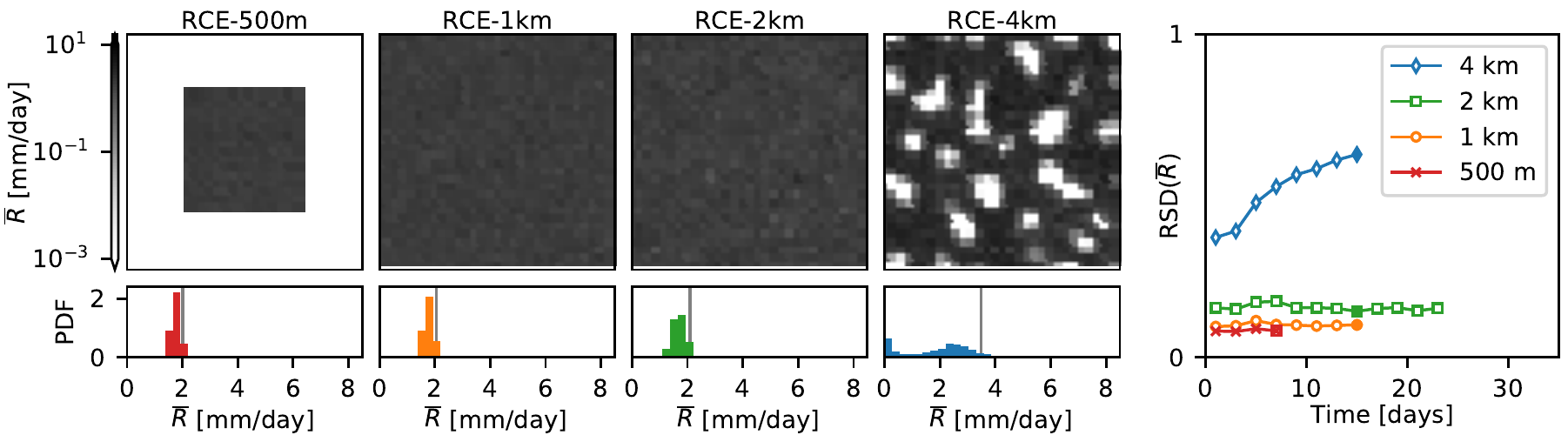}
    \put(0,25){\textbf{a}}
    \put(73,25){\textbf{b}}
    \end{overpic}
    \begin{overpic}{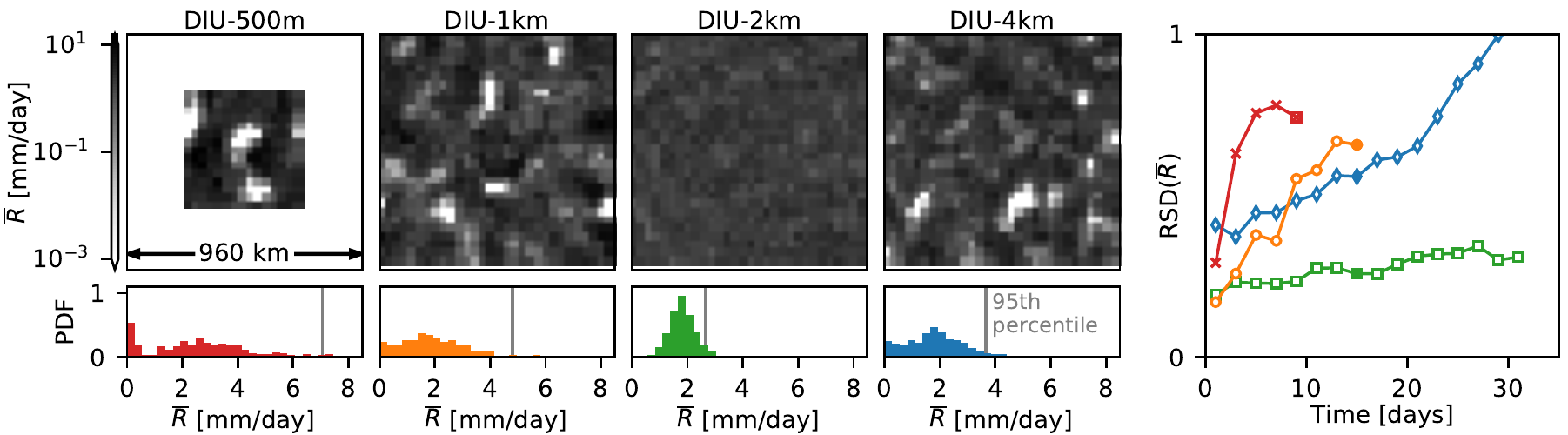}
    \put(0,25){\textbf{c}}
    \put(73,25){\textbf{d}}
    \end{overpic}
    \caption{{\bf Aggregation improved at higher resolution.} 
    \textbf{a}, Classical CSA under RCE conditions, where 
    the coarse-grained rainfall field $\overline{R}$ is shown at decreasing resolutions.
    %Surface rain intensity coarsened to grid-boxes of 32 km $times$ 32 km $times$ 48 hours.
    Note that spatial patterns only emerge at coarse resolution (4~km).
    The barplots show the histogram of $\overline{R}$ for the corresponding field in each panel.
    Grey vertical lines indicate the respective 95th percentile.
    \textbf{b}, Time series of the relative standard deviation of $\overline{R}$ for each 48-hour period. 
    \textbf{c} and \textbf{d}, Analogous to (a) and (b), but for DIU. 
    Note that strong phase separation, akin to CSA, now increasingly occurs at finer resolution. 
    Time points used in (a) and (c): RCE-500m: $t\in [6d,8d]$, DIU-500m: $t\in [8d,10d]$, all other cases: $t\in [14d0h,16d0h]$, as indicated by the solid symbols in panels (b) and (d).}
    \label{fig:resolution}
\end{figure*}

% \noindent
% {\bf Clustering strengthened by high resolution.}
\subsection*{Stronger clustering at higher resolutions}
To quantify spatial rainfall variability at scales beyond that of individual raincells, we compute the relative standard deviation (RSD) of a coarse-grained rainfall field $\overline{R}$,
where rainfall is block-averaged over 32~km $\times$ 32~km horizontally and 48 hours temporally (Details: Methods). 
The coarse-graining discounts small-scale fluctuations due to individual raincells and their cold pools, as well as any day-to-day alternation.
The 32~km spatial scale is practically relevant as it corresponds the size of large metropolitan areas.

% In RCE-4km a clear trend of increasing RSD($\overline{R}$) is seen, a feature typical of classical CSA.\cite{bretherton2005energy,wing2017convective}
RCE-4km shows a clear trend of increasing RSD($\overline{R}$), a typical feature of classical CSA\cite{bretherton2005energy,wing2017convective}.
% DIU-4km also aggregates at a rate similar to that of RCE-4km, which leads us to conclude that --- despite the diurnal cycle --- standard feedbacks of CSA\cite{emanuel2014radiative,muller2015favors, wing2017convective} are at work at this coarse resolution.
However, for horizontal resolutions of 2~km or finer, the RCE experiments show no sign of CSA (Fig.~\ref{fig:resolution}a,b) and RSD$(\overline{R})$ remains constant over time.
%constant and is slightly larger for coarser resolutions, but remains constant in time.
% This result agrees with the existing literature, which generally states that CSA is easier to achieve in RCE simulations at coarser resolution \cite{muller2015favors,wing2017convective}.
This result agrees with the existing literature, which generally states that in RCE simulations CSA is inhibited by fine resolutions \cite{muller2015favors,wing2017convective}.

In DIU, however, persistent rain-free patches are visible at 1~km resolution resulting in RSD$(\overline{R})$ rapidly increases over time.
The increase is even more pronounced in DIU-500m, despite the smaller domain size which is also often considered detrimental to CSA\cite{ muller2012detailed,jeevanjee2013convective,yanase2020new} (Fig.~\ref{fig:resolution}c,d).
In DIU-2km, RSD$(\overline{R})$ shows a weaker increase and persistent rain-free patches are all but absent (Fig.~\ref{fig:resolution}c,d). 
These observations lead us to conclude that the diurnal mechanism responsible for producing persistent dry patches is stronger with increasing horizontal resolutions.
%However, even after 30 days no persistently rain-free patches are found.
At very coarse resolution DIU-4km also aggregates at a rate similar to that of RCE-4km.
We ascribe this to known CSA feedbacks, acting despite the diurnal cycle.

%Quantitative detail on extreme precipitation can be obtained from the histograms of $\overline{R}$ (Fig.~\ref{fig:resolution}a,c), which highlight, that 
Extreme precipitation is strongly enhanced for high resolutions in DIU ({\it see} grey lines in Fig.~\ref{fig:resolution}c). 
In RCE, extremes remain comparably small at such resolutions. 
%({\it see} arrows in Fig.~\ref{fig:resolution}a) and tails decay sharply (Fig.~\ref{fig:precipitation}).
%\textcolor{red}{At the practically-relevant daily timescale, the distribution functions corresponding to DIU have exponential tails, which systematically broaden over time and with finer resolution (Fig.~\ref{fig:precipitation}). As an index for daily rainfall extremes\cite{lenderink2008increase}, the 99th rainfall percentile increases fourfold in DIU compared to RCE at 500m and 1km horizontal resolutions. % --- a finding explained by the pronounced spatial clustering in DIU.}
The 99th percentile of daily rainfall---a typical index of extreme precipitation\cite{lenderink2008increase}---increases fourfold in DIU compared to RCE at 500m and 1km horizontal resolutions, also measured at horizontal box-size of 32 km$\times$32 km
%The full distributions of daily rainfall are available in the supplementary material 
({\it Details}: Fig.~\ref{fig:precipitation}).

%{\color{red} [Romain's part starts here]}
\begin{figure*}[ht]
    \centering
    \includegraphics[width=.9\textwidth]{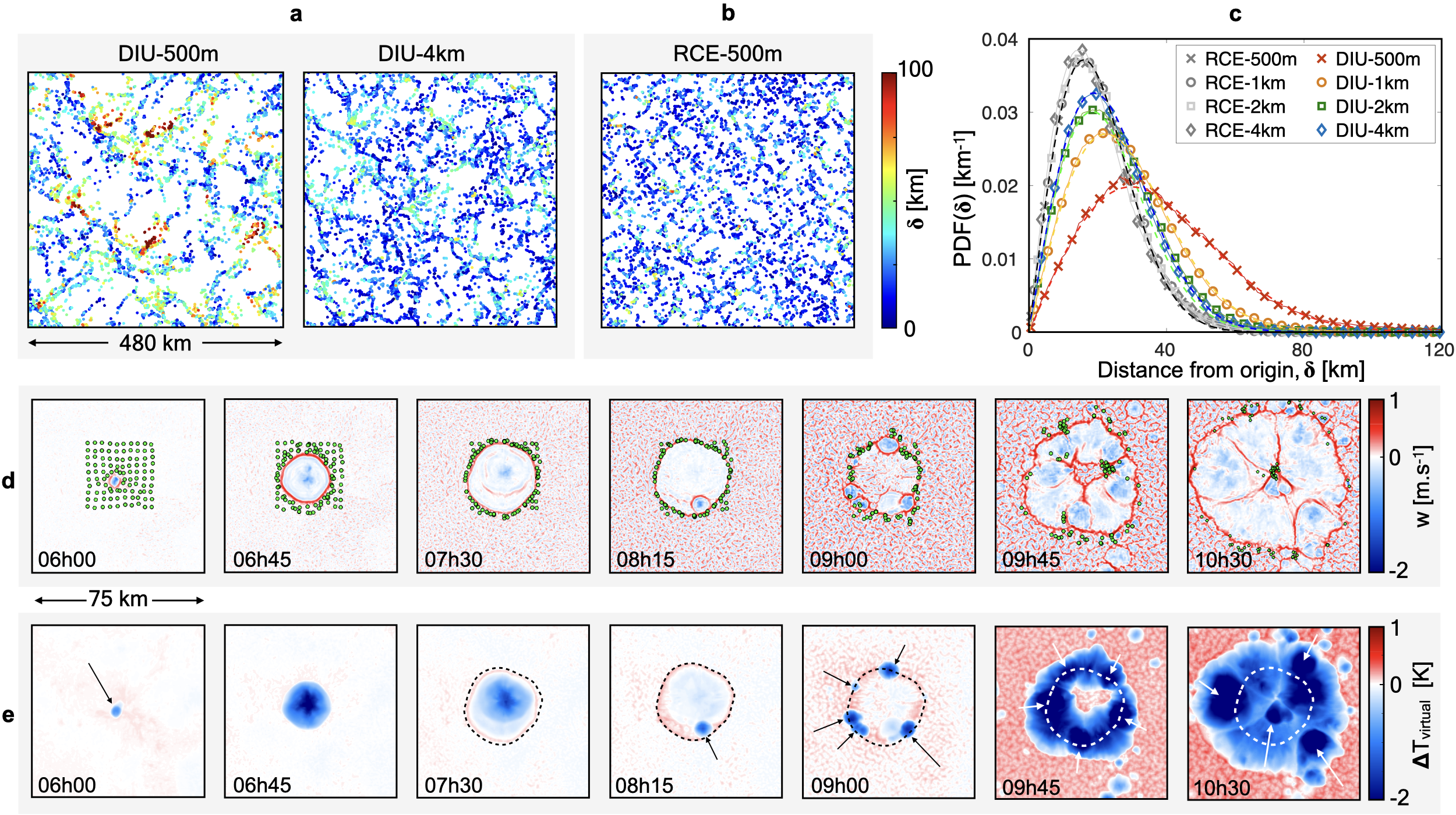}
    \caption{{\bf Low-levels circulation enhanced by resolution and diurnal cycle.}
    {\bf a}, Final state of the Lagrangian particle tracking analysis in DIU. 
    Particles are seeded along a 4~km square lattice at 1d6h, and plotted at 2d6h and coloured by the respective distance traveled, $\delta$.
    {\bf b}, Analogous to (a) but using RCE-500m. 
    All RCE cases yielded a visually similar particle field and are not presented for conciseness.
    {\bf c}, Probability distribution functions of $\delta$.
    Dashed lines corresponds to Rayleigh distribution best-fits.
    The best-fit scale parameters $\sigma$ are $[37.1, 27.2, 24.3, 23.2]$ km for DIU for $dx=[0.5, 1.0, 2.0, 4.0]$~km, respectively, whereas $\sigma=18.8\pm 0.9$~km for RCE.
    {\bf d}, Instantaneous vertical velocity fields at z=50 m during day 1 (exact times as labelled) showing the evolution of a {\emph{primo}}-CP.
    %birth, expansion and evolution.
    10 $\times$ 10 seeds, uniformly-distributed on square lattice with spacing of 4~km, are initialised at 5h to visualise the surface flow.
    {\bf e}, Analogous to (d) but for the virtual temperature anomaly field $\Delta T_{\mathrm{virtual}}$ defined as the local difference to the (z=50 m)-horizontal average.
    Arrows highlight new CPs.
    The dashed line correspond to the convergence ring of the {\emph{primo}}-CP front once it has stopped its first expansion phase ({\it Details}: Methods).
    }
    \label{fig:LPT}
\end{figure*}

%Returning to the spatial rainfall patterning (Fig.~\ref{fig:resolution}a,c), we ask 
%\noindent
\subsection*{Convective cascades and combined CPs}
Why do convective activity aggregate at high resolution for DIU, but not for RCE? 
We examine the cascade of events leading to different organisational patterns in DIU versus RCE
by mapping the low-level horizontal flow using Lagrangian particle tracking.
In DIU, seeds are spaced regularly before the onset of precipitation at 1d6h and passively advected with the horizontal flow during 24 hours ({\it Details}: Methods). 
We perform an analogous particle tracking for RCE.
In DIU-500m the final particle positions are visually clustered into stringy structures (Fig.~\ref{fig:LPT}a).
Such patterns, which we attribute to the gust fronts of combined CPs produced by MCSs, are hardly visible in the coarser-resolution simulation DIU-4km. Also, particles in DIU-500m are generally displaced much further than in DIU-4km, with large cleared spaces opening up in DIU-500m ({\it compare} panels in Fig.~\ref{fig:LPT}a). 
To quantify these differences, we compute the distance between the initial and final position of each particle, termed $\delta$. 
The set of distances for all particles yields histograms of $\delta$ (Fig.~\ref{fig:LPT}c) for all simulations.
These distributions are all remarkably well-fitted by Rayleigh functions: $\pi (\nicefrac{\delta}{2\sigma}) \exp[-\pi(\nicefrac{\delta}{2\sigma})^2]$.
The Rayleigh function describes the radial part of a two-dimensional normal distribution and is consistent with the motion of a random walker.
The diffusive length scale $\sigma$ measures the typical distance travelled. 
% The appropriateness of this fit suggests that particle motion is well-described as a random walk.
Interestingly, $\sigma$ systematically increases for DIU as model resolution is made finer.
In RCE %, where particles do not visually organise (Fig.~\ref{fig:LPT}b), 
resolution has no noticeable effect on $\sigma$, which is consistently smaller than for DIU (Fig.~\ref{fig:LPT}c, grey curves).

For DIU, the continued increase of $\sigma$ at our highest resolution (DIU-500m) suggests that processes exist, which require a mesh even finer than $500\,m$ to become fully activated.
What are these small-scale processes which have such a dramatic impact on the low-level circulation? 
Why are these processes only relevant in DIU, but not in RCE?
RCE simulations are characterised by seemingly random eruptions of convective raincells and associated spread of CPs with CP diameters of typically 10~km. 
These CPs are the main cause of horizontal winds, and they quickly transport tracers from their interior to the gust fronts.
However, the disorganised occurrences of new CPs prevent the tracers from travelling long distances effectively.

By contrast, raincells in DIU occur in diurnal bursts, where the first raincells on a given day erupt within a quiescent, but near-unstable, environment (Fig.~\ref{fig:LPT}d,e).
% We term the CP resulting from such initial raincells as \emph{primo-CP}, which, in a sense, succeeds in the competition for the most unstable locus --- and can freely expand.
We term the CP resulting from the initial raincells which succeeds in the competition for the most unstable locus as \emph{primo-CP}.
%With this competitive advantage, such primo-CPs can then freely expand. 
%By further destabilising the surrounding environment through positive vertical velocity (Fig.~\ref{fig:LPT}d) and buoyancy perturbations (Fig.~\ref{fig:LPT}e) these primo-CPs can set off a cascade of secondary raincell-CP pairs ({\it see} 8h15-9h00), which in turn instigate a tertiary population ({\it see} 10h30).
By further destabilising the surrounding environment through positive vertical velocity (Fig.~\ref{fig:LPT}d), these primo-CPs can set off a cascade of secondary raincell--CP pairs ({\it see} 8h15-9h00).
The secondary CPs instigate a tertiary population ({\it see} 10h30) and so forth.
Since such cascades of convective activity can last for several hours and span more than $\sim 100$~km we view them as emergent MCSs\cite{houze2004mesoscale}.
This process is driven by the merging of outward-running fronts into an enclosing macro-structure we refer to as \emph{combined CP}\cite{haerter2020diurnal}.
The large areas of the combined CPs result in a more persistent tracer transport (green points in Fig.~\ref{fig:LPT}d), explaining the increased mean $\delta$'s.
Similar MCS-like expansion processes occur throughout the model domain, on the same and on subsequent days.
Eventually, the interaction with other combined CPs and diurnally decreasing surface forcing halt further expansion.
After the cascade is completed, a large region remains convectively suppressed by the reduced buoyancy of the cold and dry boundary layer (Fig.~\ref{fig:LPT}d,e, 10h30).
Especially the lack of moisture is responsible for the negatively day-to-day rainfall correlations.
Importantly, two conditions have to be met to enable such cascades: 
(i) strong nocturnal cooling, ensuring quiescent conditions and thus the existence of a primo-CP; 
(ii) a mesh fine enough to resolve the convergence ring and the vertical mass fluxes at its edges --- allowing the primo-CP to transition into a combined CP. 
Note that (i) cannot be satisfied by RCE, thus hampering combined CPs.
Conversely, (ii) is likely not met by DIU-2km and DIU-4km, impeding the cascade mechanism and opening for standard CSA feedbacks to take over ({\it see} Fig. \ref{fig:2Ddivergence_allCases} for additional illustration).

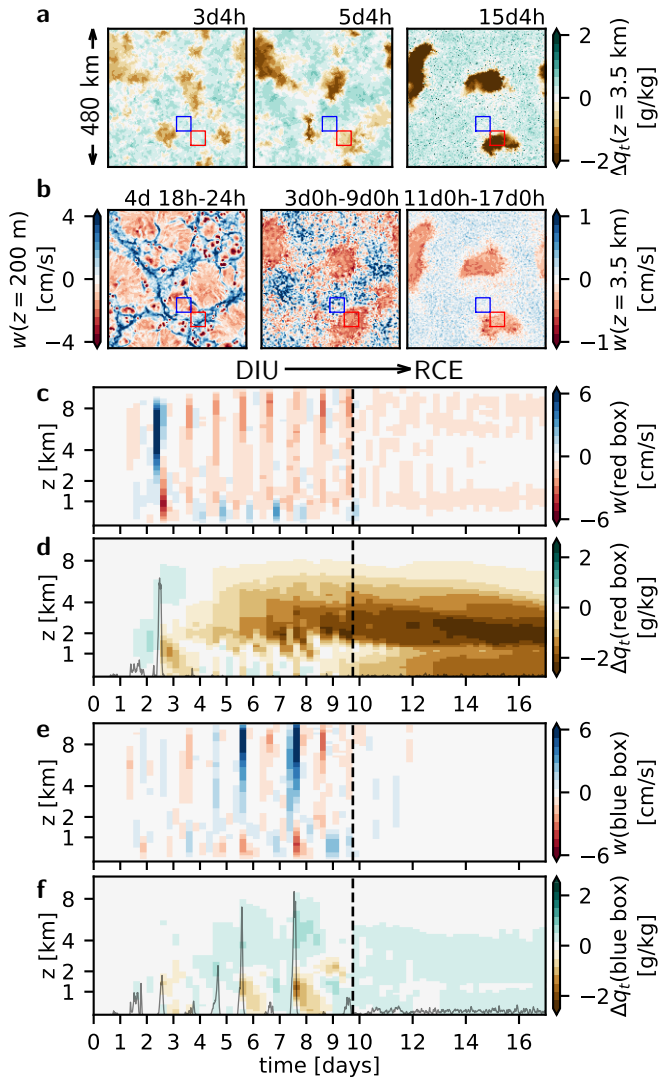
\begin{figure}[ht]
    \begin{overpic}{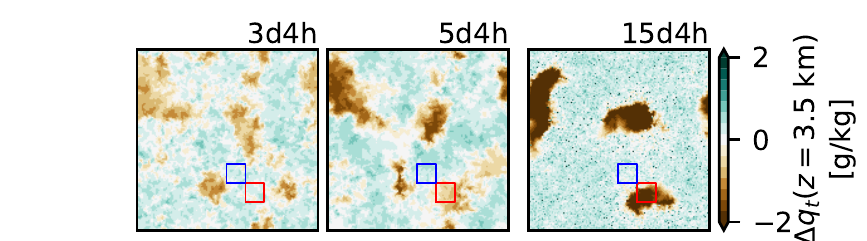}
        \put(5,23){\textbf{a}}
        % Horizontal length-scale label
        \put(13,12){\makebox(0,0){\rotatebox{90}{480 km}}}
        \linethickness{1pt}	
        \put(13.5,22){\color{black}\line(0,-1){3}}   % Top
        \put(13.5,22){\color{black}\line(0.3,-1){0.5}}
        \put(13.5,22){\color{black}\line(-0.3,-1){0.5}}
        \put(13.5,1.5){\color{black}\line(0,1){3}}     % Bottom
        \put(13.5,1.5){\color{black}\line(0.3,1){0.5}}
        \put(13.5,1.5){\color{black}\line(-0.3,1){0.5}}
    \end{overpic}
    \begin{overpic}{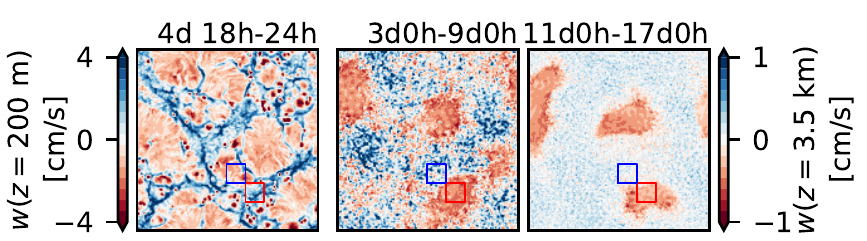}
        \put(5,24){\textbf{b}}
        % DIU->RCE label
        \linethickness{1pt}	
        \put(35,-3.5){DIU}
        \put(62,-3.5){RCE}
        \put(61.5,-2){\color{black}\line(-1,0){19}}   % Right
        \put(61.5,-2){\color{black}\line(-1,0.3){2}}
        \put(61.5,-2){\color{black}\line(-1,-0.3){2}}
    \end{overpic}\vspace{3mm}
    \begin{overpic}[trim={0 9mm 0 0},clip]{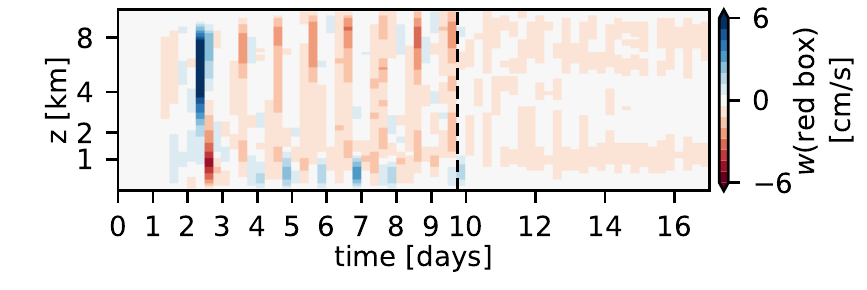}
        \put(5,20){\textbf{c}}
    \end{overpic}
    \begin{overpic}[trim={0 5mm 0 1mm},clip]{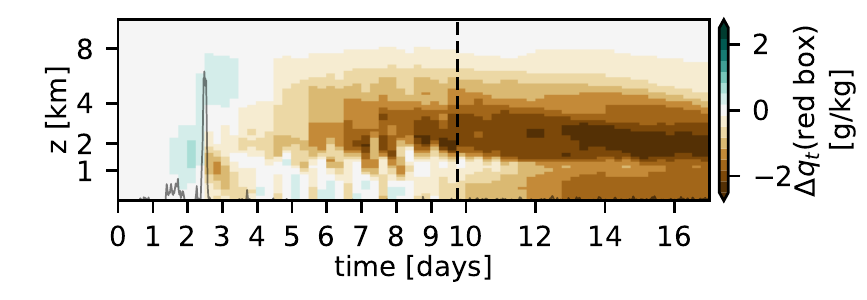}
        \put(5,24){\textbf{d}}
    \end{overpic}
    \begin{overpic}[trim={0 9mm 0 0},clip]{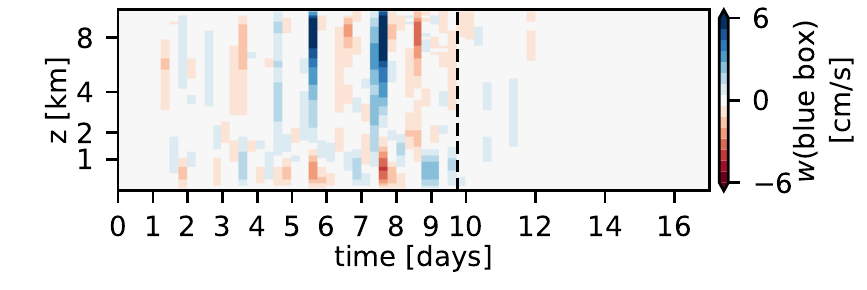}
        \put(5,20){\textbf{e}}
    \end{overpic}
    \begin{overpic}[trim={0 0 0 1mm},clip]{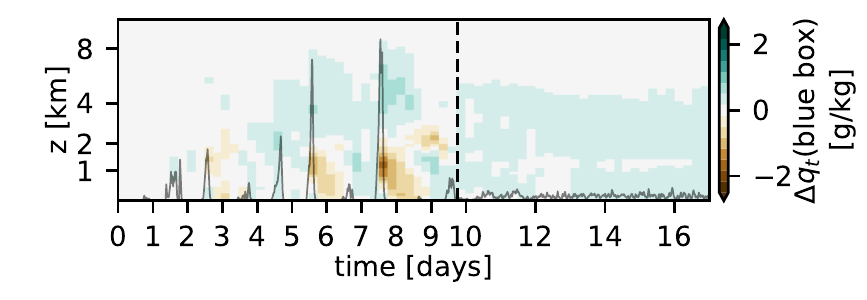}
        \put(5,29){\textbf{f}}
    \end{overpic}
\caption{\textbf{Onset of a dry patch and transition to RCE.} 
    \textbf{a}, Instantaneous nocturnal (4h) horizontal moisture anomaly at $z=3.5$~km after 3, 5, and 15 days. 
    $\Delta q_t \equiv q_t-\langle q_t\rangle$ where $\langle q_t\rangle$ is the average value at the given temporal and vertical coordinate $(t,z)$.
    \textbf{b}, Vertical velocity at $z=$200 m (left panel) and $z=$ 3.5~km (centre and right), time-averaged as noted along panels.
    \textbf{c}, Vertical velocity mean, where $w(t,x,y,z)$ was averaged horizontally within the region marked by red boxes in (a) and (b) and averaged-temporally within six-hour intervals (0h---6h, 6h---12h, 12h---18h, and 18h---24h on each day).
    \textbf{d}, Instantaneous horizontal average of total water mixing ratio, $q_t(t,z)$, within the red box at times 4h, 10h, 16h, and 22h on each day.
    \textbf{e} and \textbf{f}, Analogous to (c) and (d), but corresponding to the blue box.
    The dashed lines in panels c--f mark the transition from oscillating (DIU-500m) to constant boundary conditions (DIU2RCE-500m).
    The grey curves in panels (d) and (f) show time series of average rain intensity in the corresponding regions. In the red box (d) the peak value at $t\approx2$d$12$h is $43$ mm h$^{-1}$. 
    % Maximum rain intensity in the red quare: 42.84 mm/h
    % Maximum rain intensity in the blue quare: 53.52 mm/h
    \label{fig:zt-plots}
    }
\end{figure}

% \noindent
% {\bf Mechanism for dry patch persistence.}
\subsection*{Genesis of a persistent dry patch}
The cascade process described above brings about boundary layer dry anomalies strong enough to suppress convective activity on the following day.
But how are dry patches enabled that persist even longer, over many days?
Consider the onset of a particular persistent dry patch in DIU-500m (Fig.~\ref{fig:zt-plots}a, red square).
% The initial formation of the dry patch can be seen as the net effect resulting from the strong updraught above approximately two kilometres, substantial rainfall, and a resulting strong combined CP, which induces subsidence below (Fig.~\ref{fig:zt-plots}c).
The formation of the persistent dry patch is initiated by a strong MCS at $t\approx$2d12h.
This process leaves the free troposphere relatively moist, but the boundary layer is dried out  (Fig.~\ref{fig:zt-plots}d). % \textcolor{red}{(Gorm: mark this in the plot somehow, maybe a discrete arrow)}
On the following day, deep convection elsewhere forces pronounced subsidence within the region of interest ($t\approx$ 3d12h).
This subsidence leads to strong drying within the free troposphere,
which experiences a change from a moist to a dry anomaly within a single day (Fig.~\ref{fig:zt-plots}d).
% The resulting dry anomaly can now self-sustain, a quality we attribute to the well-known moisture-radiation feedback invoked in studies on the maintenance of traditional CSA
At this stage the resulting dry anomaly becomes self-sustaining. 
We attribute that to the well-known moisture-radiation feedback invoked in studies on the maintenance of traditional CSA
\cite{bretherton2005energy,muller2012detailed,muller2015favors,yanase2020new}:
the dry free troposphere (Fig.~\ref{fig:zt-plots}a,c) gives rise to increased long-wave cooling ({\it compare}: Fig.~\ref{fig:diu2rce_SI}i---l), which in turn must be compensated by general subsidence heating (Fig.~\ref{fig:zt-plots}b). 
Subsidence further amplifies the drying and prevents deep convective activity. %  within the persistent dry 

In the boundary layer a circulation is driven by the CP outflow from surrounding deep convective activity (Fig.~\ref{fig:zt-plots}b,4d18h-24h) and results in significant evening updraughts below $z\approx 1$~km between days four and eight (Fig.~\ref{fig:zt-plots}c).
Such nocturnal low-level updraughts do not initiate new convection because the atmosphere is already stabilised at this time of day.

% Outside the dry patches, the dynamics changes from day to day, exemplified by the blue-boxed area in Fig.~\ref{fig:zt-plots}a,b and Fig.~\ref{fig:zt-plots}e,f.
Outside the dry patches, the dynamics switches from day to day between two modes (Fig.~\ref{fig:zt-plots}a,b blue-boxed and e,f):
on some days (eg. 6th, and 8th), mid-day convective activity results in heavy rainfall, increased free-tropospheric moisture, and significant drying in the boundary layer.
On other days (eg. 7th and 9th) the dynamics resembles that within the persistent dry patch, with net mid-day subsidence followed by low-level nocturnal updraughts.
However, the mid-day subsidence is weaker than in the dry patch, and the free troposphere does not dry below the domain average, so convection is no longer inhibited once the boundary layer has re-moistened.
Rapid free-tropospheric drying appears to be a pivotal step in persistent dry patches formation.

% text for Fig. 5
Periodic surface temperature forcing can induce persistent dry patches,
but can these dry patches prevail when the periodic forcing is removed?
To answer this, we extend DIU-500m, starting with its state at 9d18h but replace the DIU with the RCE forcing, that is constant surface temperature and insolation. %.
As might be expected, the DIU-to-RCE switch leads to a gradual relaxation of domain mean atmospheric temperature and rainfall to a near-constant steady state (Fig.~\ref{fig:domain_mean_timeseries}).
Inspecting total moisture at $z=3.5$~km (Fig.~\ref{fig:zt-plots}a and Fig.~\ref{fig:diu2rce_SI}a--h), the spatial partitioning into moist and dry patches is clearly preserved and dry patches sharpening significantly, especially in the boundary layer %, with horizontal moisture distributions becoming increasingly bi-modal 
(Fig.~\ref{fig:diu2rce_SI}d,h). 
Hence, the transition to RCE causes an alignment of dry patches throughout the vertical column---a feature consistent with classical CSA.
%Interested readers can find more information about the DIU2RCE extension in the supplementary material (available online).
The classical CSA trademarks of increased long-wave cooling and the absence of rainfall over dry areas\cite{bretherton2005energy}
%Indeed, the state obtained after relaxation in DIU2RCE shows that outgoing long-wave radiation is increased for the dry patches 
are recovered (Fig.~\ref{fig:diu2rce_SI}i---k, m---o).
%and surface precipitation is all but absent (Fig.~\ref{fig:diu2rce_SI}n---p). 
We conclude that temporary convective organisation under DIU can lead to persistent CSA-like organisation, even after the periodic forcing is removed.

% \noindent
% {\bf Implications for tropical climate modelling.}
\subsection*{Implications for tropical climate modelling}
The numerical modelling of large-scale thunderstorm clustering was singled out as one of the fundamental questions relevant to the global climate \cite{bony2015clouds}.
One prominent explanation for such clustering was CSA,\cite{bretherton2005energy}
but CSA has been found difficult to achieve at high model resolution when CPs are vigorous \cite{muller2012detailed,jeevanjee2013convective,muller2015favors}.
%Such "cloud clumping" was previously described as "gregarious,"\cite{mapes1993gregarious} where mesoscale groups of clouds are, perhaps startlingly, tied together in even larger "super-clusters." 
Our results show that diurnal convective activity spontaneously organises into large-scale spatio-temporal patterns of MCSs and persistently dry patches, especially when model resolution is high, that is, within the limit of greatest practical relevance.
%Our results show that clustering into a coherent pattern of MCSs and persistently dry patches emerges spontaneously, especially when model resolution is high, hence within the limit of greatest practical relevance.
% Our results show that clustering into an entangled super-cluster of MCS can self-emerge especially when model resolution is high --- hence, within the limit of greatest practical relevance.
Previously, conceptual work attributed the onset of CSA to a linear instability in the free troposphere \cite{emanuel2014radiative}.
% CSA was however found difficult to achieve at high model resolution or when CPs are vigorous \cite{muller2012detailed,jeevanjee2013convective,muller2015favors}.
In contrast, the onset of persistent dry patches in our simulations is a highly non-linear boundary layer process set off by strongly-correlated CP dynamics. 
%Yet, the mechanisms for stabilisation of these disturbances again seem in line with known feedbacks \cite{muller2015favors}. 
%From this population persistent dry patches aris, with striking similarity to those in CSA.
%In RCE experiments, the onset of CSA is controlled by a competition a positive moisture-radiation feedback in the free troposphere and a negative feedback in the boundary layer driven by CPs.
%Our results show that in the presence of diurnal surface temperature oscillations CPs can form merge together forming combined CPs.
%The resulting combined CP diameters can be an order of magnitude larger than for individual CPs.
The role of large combined CPs can be seen as inverted, because they can suppress convective activity for long enough to push the troposphere beyond the tipping point at which dry patches become self-perpetuating.

%We have also demonstrated that the diurnal cycle is only necessary for triggering the onset of dry spots.
Once formed, dry patches can persist and intensify under temporally constant boundary conditions, with potential consequences for the interface between tropical continents and oceans, such as that between tropical Africa and the Atlantic Ocean: % our results suggest 
a persistent super-cluster, generated over the continent, could be advected over the sea, where the clustering further intensifies, even under constant surface temperatures.
%It should be further explored, which influence seasons and time of day have on the ability of clusters to persist and grow when crossing the Western continental.
%The clustering in our simulations is facilitated at finer horizontal model resolution, and the 5K surface temperature amplitude used in our simulations is sufficient to instigate the mesoscale convective clustering we report.  
%Indications for $\mathcal{O}(1\,K)$ oceanic surface temperature oscillations exist, when surface wind is persistently weak and insolation strong. 
In addition, oceanic surface temperature amplitudes as large as two \cite{weller1996surface,johnson1999trimodal} to five kelvin \cite{kawai2007diurnal} have been observed, and oscillations can %so-called diurnal warm layers 
persist for up to five days\cite{bellenger2009analysis}, with a measurable impact on atmospheric properties. \cite{kawai2007diurnal,bellenger2009analysis,bellenger2010role}
%As fine resolution and larger diurnal surface temperature amplitudes both promote persistent clustering, it should therefore be explored, if clustering prevails when the amplitude is reduced but resolution further increased. 
%without loosing the effect of mesoscale clustering, and if this threshold amplitude might even reduce to smaller values, as model resolution is further increased.
 
Regarding extreme convective precipitation, consistently shown to intensify strongly under daily-mean surface temperature increases \cite{lenderink2008increase,berg2013strong,moseley2016intensification}, our findings offer a complementary view: 
extremes can increase, even if the daily-mean temperature is fixed.
Our results suggest that convective precipitation extremes are also influenced by the forcing temperature diurnal range and by the state of self-organisation induced by the diurnal cycle.

%Simple models of tropical cloud clustering date back at least to \citeauthor{randall1980stochastic} (\citeyear{randall1980stochastic}) and the fundamental organisational mechanisms can inspire fascinating parallels between disciplines \cite{hubbell1979tree}.
%The current numerical experiments encourage  "toy models," that capture the clumping leading to MCS formation, the larger-scale communication between MCSs, as well as the emergence of persistent dry spots.
%Whereas sub-kilometre resolution global cloud resolving simulations may ultimately capture such emergence under further growing computing capabilities, fundamental understanding could be strengthened by further efforts in conceptual modelling.

\clearpage
\section*{Materials and Methods}\label{sec:materials}
\noindent
{\bf Large-eddy model, boundary, and initial conditions.}
To simulate the convective atmosphere, we employ the University of California, Los Angeles (UCLA) Large Eddy Simulator (LES) with sub-grid scale turbulence parametrised after Smagorinsky \cite{smagorinsky1963general}.
The Coriolis force and the mean wind are both set to zero.
Radiation effects are incorporated using  a delta four-stream scheme \cite{pincus2009monte} and a two-moment cloud microphysics scheme \cite{stevens2005evaluation}. 
Rain evaporation depends on ambient relative humidity and the mean and spread of hydrometeor radii\cite{seifert2006two}.
Long and shortwave radiation interacts with the atmosphere including clouds, but it does not impact the surface temperature, which is prescribed and spatially homogeneous.
The prescribed surface temperature $T_s(t)$ is spatially homogeneous but oscillates temporally as 
\begin{eqnarray}
    T_s(t)=\overline{T_{s}}-T_{a} \cos{(2\pi\;t/t_0)}\;,
    \label{eq:T_s_standard}
\end{eqnarray}
\noindent
where $\overline{T_{s}}=298\;K$, $t_0=24\;h$ is the period of the simulated model day, $\overline{T_{s}}$ is the temporal average and $T_a$ the amplitude of $T_s(t)$.
For the simulations "DIU" $T_a=5\,K$ is chosen, whereas for "RCE" $T_a=0$.
Insolation $S(t)$ is taken as spatially homogeneous for all simulations.
For the simulations "DIU" the insolation cycle $S(t)$ oscillates temporally with an amplitude typical for the equator. 
For the "RCE" simulations, both $T_s(t)$ and $S(t)$ are set constant to their respective temporal averages, that is, $T_s(t)=\overline{T_{s}}=298\;K$ and $S(t)=\overline{S}=445\;W\,m^{-2}$.

\noindent
{\bf Surface latent and sensible heat fluxes.}
Surface heat fluxes are computed interactively by standard bulk formulae and increase with the vertical temperature and humidity gradients as well as horizontal wind speed.
Horizontal surface wind speed is approximated through Monin-Obukhov similarity theory \cite{stull2012introduction}.
Our simulations use a simple parametrisation of a homogeneous, flat land surface, by assuming surface latent heat fluxes to be reduced to 70 percent of those obtained for a saturated (sea) surface.
For the DIU experiments, mean surface latent and sensible heat fluxes are $LHF\approx 57\;W/m^2$ and $SHF\approx 18\;W/m^2$, respectively, yielding a Bowen ratio of $B\approx .30$, realistic for forested land. 
Initial temperature and humidity are taken from observed profiles that potentially represent convective conditions\cite{moseley2016intensification}, but quickly self-organise during the initial spin-up.
The initial temperature field is perturbed by small uncorrelated noise terms to allow the simulation to break complete spatial symmetry. 
The spin-up manifests itself in "DIU" by relatively weak precipitation during the first model day, but relatively strong precipitation during the second. 
From the third day on, precipitation diurnal cycles are found fairly repetitive ({\it compare}: Fig.~\ref{fig:DIU-1km}k).
Hence, over time, the system eventually establishes a self-consistent vertical temperature and moisture profile.

\noindent
{\bf Sensitivity studies.} % As we prescribe surface temperature, the prescribed shortwave diurnal forcing (insolation) does not modify the surface conditions, and therefore has a minor impact on the results obtained here compared to the prescribed surface temperature, which dominates the signal. 
As we prescribe surface temperature, its value is unaffected by the shortwave diurnal forcing (insolation). 
Modifications to insolation therefore have only a minor impact on the results:
shifting the phase of prescribed surface temperature relative to that of insolation or setting insolation constant, did not yield any qualitative changes to the diurnal clustering. 
In addition, taking the surface as a water surface had only minor effects on the diurnal clustering.
These and further sensitivities were previously explored in detail\cite{haerter2020diurnal}.
Additionally, we explored changes to initial conditions, by restarting a simulation similar to DIU-1km, but using the horizontal means of temperature and water vapour, $T(z)$ and $q_v(z)$, respectively, obtained after several days of a simulation analogous to RCE-1km, as initial condition.
The diurnal clustering and gradual emergence of dry patches was unaffected.

\noindent
{\bf Model grid, dynamics, and output.}
The anelastic equations of motion are integrated on a regular horizontal domain with varying horizontal grid spacing $dx$ and laterally periodic boundary conditions (Tab. \ref{tab:experiments}). 
Vertically, the model resolution is stretched, with $100$~m below $1$~km, $200$~m near $6$~km, and $400$~m in the upper layers.  
A sponge layer is implemented between $12.3$~km and the model top, which is located at $16.5$~km.
Horizontal resolution $dx$, domain size, and output timestep $\Delta t_{out}$ vary (Tab.~\ref{tab:experiments}).
%No large scale forcing was imposed, ensuring that the only driving force for convection was buoyancy and the forced lifting through cold pool interaction.
%For all two and three-dimensional model variables, the output timestep was set to $\Delta t_{out}=20\;min$. 
At each output timestep, instantaneous surface precipitation intensity, as well as instantaneous horizontal fields of velocity and thermodynamic variables at various vertical levels are recorded.
Three-dimensional thermodynamic output data are recorded instantaneously at UTC 4, 10, 16, and 22, whereas three-dimensional velocities are recorded as time averages between UTC 0---6, 6---12, 12---18 and 18---24.
Additionally, at 30-second and five-minute intervals, respectively, spatially as well as horizontally averaged time series were extracted from the numerical experiments.

\begin{table*}[ht]
\centering
\begin{tabular}{llllll}
    %\rowcolor[HTML]{ECF4FF} 
    Experiment & Surface temperature & Horizontal resolution & Domain size & Simulation & Output timestep\\
    name & amplitude, $T_a$ [K] & $dx$~[km] & $L$~[km] & period [days] & $\Delta t_{out}$~[min]\\
    \hline
    DIU-500m & $5$ & $0.5$ & $480$ & $0$---$10$ & 15\\
    RCE-500m & $5$ & $0.5$ & $480$ & $0$---$8$ & 20\\
    DIU2RCE-500m & $0$ & $0.5$ & $480$ & $9.75$---$17$ & 15\\
    DIU-1km & $5$ & $1$ & $960$ & $0$---$16$ & 20\\
    DIU2RCE-1km & $0$ & $1$ & $960$ & $15.75$---$24$ & 20\\
    RCE-1km & $0$ & $1$ & $960$ & $0$---$16$ & 20\\
    DIU-2km & $5$ & $2$ & $960$ & $0$---$24$ & 20\\
    RCE-2km & $0$ & $2$ & $960$ & $0$---$20$ & 20\\
    DIU-4km & $5$ & $4$ & $960$ & $0$---$42$ & 20\\
    RCE-4km & $0$ & $4$ & $960$ & $0$---$20$ & 20\\
    \hline
\end{tabular}
\caption{{\bf Summary of numerical experiments.}
The term "DIU" is used to indicate simulations with diurnally oscillating surface temperature $T_s(t)$ and insolation $S(t)$, whereas in "RCE" both $T_s(t)$ and $S(t)$ are held constant.
The term "DIU2RCE" means that "RCE" boundary conditions are applied as a continuation of "DIU" for the respective previous period --- such as DIU2RCE-500m, which is initialised with the three-dimensional atmospheric state after 9.75 days. 
The experiment names further include the respective horizontal model resolution.
}
\label{tab:experiments}
\end{table*}

%\noindent
%{\bf Raincell tracking.}
%The surface rain intensity space-time field is stored as a three-dimensional array $R(t,x,y)$.
%We define a raincell as a connected cluster of entries with a rain-intensity above 1 mm/h. 
%Connected clusters are defines in the following way:
%The array is treated as a network where each entry is represented by a node.
%Each node has six links, the two nearest neighbours in each direction ($t$, $x$, and $y$).
%A connected cluster is a set of nodes where there's a path from any one node to any other, along the network links, without touching any nodes outside the connected cluster.
%Practically we use the function "find$\_$objects" from the python library "scipy.ndimage".
%It should be mentioned, that our algorithm does not respect the periodic boundary condition. 
%Some raincell are therefore artificially separated. This happens only very rarely (with a domain size 960 km$times$960 km and a typical raincell diameter of 5 km, the probability that a raincell crosses the boundary is less than $10/960\approx 1\%$ ). 

\noindent
{\bf Temporal correlation.}
We define the 24-hour lag correlation $C_{24h}(q_t;t,z)$ used in Figs~\ref{fig:DIU-1km} and ~\ref{fig:DIU-500m} as the pixel-by-pixel Pearson correlation between time $t$ and $t+24h$, of the horizontal moisture distribution at the vertical level $z$:
\begin{eqnarray}
    C_{24h}(q_t;t,z) %&\equiv&
    %PC_{xy}\Big(q_t(t,x,y,z),q_t(t+24h,x,y,z)\Big) \nonumber \\
    \equiv
    \sum_{i,j=1}^{N}
    \tilde{q_t}(t,x_i,y_j,z) \,
    \tilde{q_t}(t+24h,x_i,y_j,z)
\end{eqnarray}
where $N=L / dx$ is the number of grid-boxes along the domain side length ({\it see} Tab.~\ref{tab:experiments}). 
The relative spatial anomalies of $q_t$ at time $t$ are defined as  
$\tilde{q_t}(t,x,y,z)\equiv  \Delta q_t(t,x,y,z)/ \sigma_{q_t}(t,z)$, where 
$\Delta q_t(t,x,y,z)\equiv q_t(t,x,y,z)-\langle q_t\rangle(t,z)$ is the absolute spatial anomaly of $q_t$ and %and $\langle\tilde{q_t}(t,x,y,z)\equiv \big(q_t(t,x,y,z)-\langle q_t\rangle(t,z) \big) / \sigma_{q_t}(t,z)$, where 
$\langle q_t\rangle(t,z)$ its horizontal average at time $t$ and vertical level $z$,
% $$  \langle q_t\rangle(t,z) \equiv 
%     \frac{1}{L_x L_y}\int_{x=0}^{L_x}\int_{y=0}^{L_y} q_t(t,x,y,z)\,dxdy
% $$
$$  \langle q_t\rangle(t,z) \equiv 
    \frac{1}{N^2}\sum_{i=1}^N\sum_{j=1}^N q_t(t,x_i,y_j,z)
$$
and $\sigma_{q_t} ^ 2$ is the horizontal variance 
% $$  \sigma_{q_t} ^ 2 (t,z)\equiv 
%     \frac{1}{L_x L_y}\int_{x=0}^{L_x}\int_{y=0}^{L_y} \big(q_t(t,x,y,z) - \langle q_t\rangle(t,z)\big)^2 \,dxdy\,.
% $$
% $$  \langle q_t\rangle(t,z) \equiv 
%     \frac{1}{L_x L_y}\int_{x=0}^{L_x}\int_{y=0}^{L_y} q_t(t,x,y,z)\,dxdy
% $$
$$  \sigma_{q_t} ^ 2 \equiv 
    \frac{1}{N^2}\sum_{i=1}^N\sum_{j=1}^N 
    \big(q_t(t,x_i,y_j,z)-\langle q_t\rangle(t,z)\big)^2\,.
$$
Note that, by definition, $C_{24h}(q_t;t,z)$ is bounded to lie between $-1$ and $+1$.
We compare horizontal fields of $q_t(t,x,y,z)$ for various values of height $z$ at $t$ chosen to represent 4h of each given day. 
At this time of day the atmosphere is generally stably stratified, convective activity is at a minimum and the moisture field is maximally smooth. 
This is an advantage because we are interested in the large scale structures, and not the precise locations of individual raincells, that typically measure only few kilometres in diameter.

\noindent
{\bf Coarse-graining procedure.}
Coarse-grained rain intensity fields, termed $\overline{R}$, are used in Fig.~\ref{fig:resolution} and \ref{fig:precipitation} to compute the relative standard deviation RSD($\overline{R}$), that is, the standard deviation divided by the mean, as well as the exceedence probability of daily precipitation intensity, respectively. 
%In Fig.~\ref{fig:resolution} we measure the persistent convective aggregation by the relative standard deviation (RSD) of the coarsened rain intensity field $\overline{R}$. 
$\overline{R}(k,l,m)$ is a three-dimensional array where each element represent a space-time cube of horizontal interval of length $s$ and temporal interval of duration $\tau$, that is, a cube of volume $s\times s\times \tau$. 
Hence, 
% It could be even more explicit by writing something like:
$$ \overline{R}(k,l,m) \equiv
    \int_{k \tau}^{(k+1) \tau} \hspace{-5mm} dt
    \int_{l s}^{(l+1) s} \hspace{-5mm} dx
    \int_{m s}^{(m+1) s} \hspace{-5mm} dy
    \, \, R(t,x,y) \, ,
$$
where $R$ is the model output instantaneous rainfall intensity (Tab.~\ref{tab:experiments}).
We choose $s$=32~km spatially. 
Temporally, Fig.~\ref{fig:resolution} uses $\tau=$48 h and Fig.~\ref{fig:precipitation} uses $\tau=$24 h.
The interval $s$=32~km is a compromise between being significantly larger than typical individual deep convective rain events yet small compared to the system size.
The interval $\tau$=48 h in Fig.~\ref{fig:resolution} is chosen to emphasise persistent structures and discount the day-to-day anti-correlated, high intensity mesoscale rain clusters.
The interval $\tau$=24 h in Fig.~\ref{fig:precipitation} is chosen to capture the natural timescale of one day and make contact to usual extreme event statistics.

\noindent
{\bf Lagrangian particle tracking.}
The particle tracking used in Fig.~\ref{fig:LPT} works in the following way:
we distribute a set of seeds over the lowest horizontal level ($z$=50 m) on the morning of the second day (1d6h). The set forms a squared lattice with one seed placed every 4~km. The particles are then transported over a 24h-period using the horizontal velocity solution and a trapezoidal method with a 15-minute timestep.

We choose to analyse the second simulation day because:
(i) The horizontal morning moisture distribution is increasingly clustered from day to day. Thus earlier days a preferable for disentangling the dynamical effects of cold pools from the thermodynamic preconditioning.
(ii) On the first day the rainfall is extremely sparse due to the spin-up from the initial condition. 
% % For the particle tracking used in Fig.~\ref{fig:LPT}, for each of the simulations we first distribute a set of seeds on a 4 km $\times$ 4 km square grid over the lowest horizontal level ($z$=50 m) on the morning of the second day (1d6h). 
% This specific time is chosen because: (i) persistent dry spots have not formed yet on the second day; (ii) the domain-wide diurnally-driven organisation is already ongoing during the second day (Fig. \ref{fig:resolution}d, red curve), with a steep increase in $RSD(\overline{R})$ already during the first few days; and (iii) a 6h morning seeding corresponds to a time of minimum convective activity following the nocturnal cooling, which leads to overall stratification. 
% Therefore, starting the tracking at this particular time allows to accurately capture the onset of convection on the second day. 
Repeating the analysis on a later days gives comparable results.

In the DIU-experiments we can seed the particles at early morning when there is close to zero convective activity following the nocturnal cooling.
This allows us to accurately capture the diurnal motion from onset of convection till the end of the last cold pools.
In the RCE-experiments there are no such silent periods, so we have to pick arbitrary beginning and end times for the tracers.

An animation of this process is presented in a supplemental video file.

\pagebreak

\section*{Corresponding Author}
Jan O. Haerter, haerter@nbi.ku.dk

\section*{Acknowledgements}
The authors gratefully acknowledges funding from the Villum Foundation (grant no. 13168). 
J. O. H. acknowledges funding from the European Research Council under the
European Union's Horizon 2020 Research and Innovation programme (grant no. 771859) and the Novo Nordisk Foundation Interdisciplinary Synergy Program (grant no. NNF19OC0057374).
The authors gratefully acknowledge high-performance computing resources from the German Climate Computing Center (DKRZ) as well as the Danish Climate Computing Center (DC3).

\section*{Author Contributions}
G. G. J. and J. O. H. jointly conceived the project idea and key findings.
G. G. J. performed data analysis and produced Figs \ref{fig:DIU-1km}, \ref{fig:resolution}, \ref{fig:zt-plots}, \ref{fig:DIU-500m} and \ref{fig:precipitation}.
J. O. H. performed the numerical simulations, performed data analysis and produced Figs~\ref{fig:domain_mean_timeseries} and \ref{fig:diu2rce_SI}.
R. F. performed data analysis and produced Figs~\ref{fig:LPT} and \ref{fig:2Ddivergence_allCases}.
All authors jointly wrote and edited the manuscript.

\section*{Competing Interest Statement}
The authors declare that they do not have competing interests.
%Acknowledgements (optional), Author Contributions, Competing Interests statement

\clearpage
\newpage
\pagebreak
\bibliography{references}

\newcommand{\noop}[1]{}
\begin{thebibliography}{10}
\expandafter\ifx\csname url\endcsname\relax
  \def\url#1{\texttt{#1}}\fi
\expandafter\ifx\csname urlprefix\endcsname\relax\def\urlprefix{URL }\fi
\providecommand{\bibinfo}[2]{#2}
\providecommand{\eprint}[2][]{\url{#2}}

\bibitem{held1993radiative}
\bibinfo{author}{Held, I.~M.}, \bibinfo{author}{Hemler, R.~S.} \&
  \bibinfo{author}{Ramaswamy, V.}
\newblock \bibinfo{title}{Radiative-convective equilibrium with explicit
  two-dimensional moist convection}.
\newblock \emph{\bibinfo{journal}{Journal of the Atmospheric Sciences}}
  \textbf{\bibinfo{volume}{50}}, \bibinfo{pages}{3909--3927}
  (\bibinfo{year}{1993}).

\bibitem{tompkins1998radiative}
\bibinfo{author}{Tompkins, A.~M.} \& \bibinfo{author}{Craig, G.~C.}
\newblock \bibinfo{title}{Radiative--convective equilibrium in a
  three-dimensional cloud-ensemble model}.
\newblock \emph{\bibinfo{journal}{Quarterly Journal of the Royal Meteorological
  Society}} \textbf{\bibinfo{volume}{124}}, \bibinfo{pages}{2073--2097}
  (\bibinfo{year}{1998}).

\bibitem{bretherton2005energy}
\bibinfo{author}{Bretherton, C.~S.}, \bibinfo{author}{Blossey, P.~N.} \&
  \bibinfo{author}{Khairoutdinov, M.}
\newblock \bibinfo{title}{An energy-balance analysis of deep convective
  self-aggregation above uniform {SST}}.
\newblock \emph{\bibinfo{journal}{Journal of the {A}tmospheric {S}ciences}}
  \textbf{\bibinfo{volume}{62}}, \bibinfo{pages}{4273--4292}
  (\bibinfo{year}{2005}).

\bibitem{wing2017convective}
\bibinfo{author}{Wing, A.~A.}, \bibinfo{author}{Emanuel, K.},
  \bibinfo{author}{Holloway, C.~E.} \& \bibinfo{author}{Muller, C.}
\newblock \bibinfo{title}{Convective self-aggregation in numerical simulations:
  a review}.
\newblock \emph{\bibinfo{journal}{Surveys in Geophysics}}
  \textbf{\bibinfo{volume}{38}}, \bibinfo{pages}{1173--1197}
  (\bibinfo{year}{2017}).

\bibitem{zhang2005madden}
\bibinfo{author}{Zhang, C.}
\newblock \bibinfo{title}{Madden-{J}ulian {O}scillation}.
\newblock \emph{\bibinfo{journal}{Reviews of Geophysics}}
  \textbf{\bibinfo{volume}{43}} (\bibinfo{year}{2005}).

\bibitem{emanuel2018100}
\bibinfo{author}{Emanuel, K.}
\newblock \bibinfo{title}{100 years of progress in tropical cyclone research}.
\newblock \emph{\bibinfo{journal}{Meteorological Monographs}}
  \textbf{\bibinfo{volume}{59}}, \bibinfo{pages}{15--1} (\bibinfo{year}{2018}).

\bibitem{muller2012detailed}
\bibinfo{author}{Muller, C.~J.} \& \bibinfo{author}{Held, I.~M.}
\newblock \bibinfo{title}{Detailed investigation of the self-aggregation of
  convection in cloud-resolving simulations}.
\newblock \emph{\bibinfo{journal}{Journal of the Atmospheric Sciences}}
  \textbf{\bibinfo{volume}{69}}, \bibinfo{pages}{2551--2565}
  (\bibinfo{year}{2012}).

\bibitem{emanuel2014radiative}
\bibinfo{author}{Emanuel, K.}, \bibinfo{author}{Wing, A.~A.} \&
  \bibinfo{author}{Vincent, E.~M.}
\newblock \bibinfo{title}{Radiative-convective instability}.
\newblock \emph{\bibinfo{journal}{Journal of Advances in Modeling Earth
  Systems}} \textbf{\bibinfo{volume}{6}}, \bibinfo{pages}{75--90}
  (\bibinfo{year}{2014}).

\bibitem{muller2015favors}
\bibinfo{author}{Muller, C.} \& \bibinfo{author}{Bony, S.}
\newblock \bibinfo{title}{What favors convective aggregation and why?}
\newblock \emph{\bibinfo{journal}{Geophysical Research Letters}}
  \textbf{\bibinfo{volume}{42}}, \bibinfo{pages}{5626--5634}
  (\bibinfo{year}{2015}).

\bibitem{coppin2015physical}
\bibinfo{author}{Coppin, D.} \& \bibinfo{author}{Bony, S.}
\newblock \bibinfo{title}{Physical mechanisms controlling the initiation of
  convective self-aggregation in a general circulation model}.
\newblock \emph{\bibinfo{journal}{Journal of Advances in Modeling Earth
  Systems}} \textbf{\bibinfo{volume}{7}}, \bibinfo{pages}{2060--2078}
  (\bibinfo{year}{2015}).

\bibitem{hohenegger2016coupled}
\bibinfo{author}{Hohenegger, C.} \& \bibinfo{author}{Stevens, B.}
\newblock \bibinfo{title}{Coupled radiative convective equilibrium simulations
  with explicit and parameterized convection}.
\newblock \emph{\bibinfo{journal}{Journal of Advances in Modeling Earth
  Systems}} \textbf{\bibinfo{volume}{8}}, \bibinfo{pages}{1468--1482}
  (\bibinfo{year}{2016}).

\bibitem{craig2013coarsening}
\bibinfo{author}{Craig, G.} \& \bibinfo{author}{Mack, J.}
\newblock \bibinfo{title}{A coarsening model for self-organization of tropical
  convection}.
\newblock \emph{\bibinfo{journal}{Journal of Geophysical Research:
  Atmospheres}} \textbf{\bibinfo{volume}{118}}, \bibinfo{pages}{8761--8769}
  (\bibinfo{year}{2013}).

\bibitem{holloway2016sensitivity}
\bibinfo{author}{Holloway, C.~E.} \& \bibinfo{author}{Woolnough, S.~J.}
\newblock \bibinfo{title}{The sensitivity of convective aggregation to diabatic
  processes in idealized radiative-convective equilibrium simulations}.
\newblock \emph{\bibinfo{journal}{Journal of Advances in Modeling Earth
  Systems}} \textbf{\bibinfo{volume}{8}}, \bibinfo{pages}{166--195}
  (\bibinfo{year}{2016}).

\bibitem{jeevanjee2013convective}
\bibinfo{author}{Jeevanjee, N.} \& \bibinfo{author}{Romps, D.~M.}
\newblock \bibinfo{title}{Convective self-aggregation, cold pools, and domain
  size}.
\newblock \emph{\bibinfo{journal}{Geophysical Research Letters}}
  \textbf{\bibinfo{volume}{40}}, \bibinfo{pages}{994--998}
  (\bibinfo{year}{2013}).

\bibitem{boye2019self}
\bibinfo{author}{Boye~Nissen, S.} \& \bibinfo{author}{Haerter, J.~O.}
\newblock \bibinfo{title}{Self-aggregation conceptualized by cold pool
  organization}.
\newblock \emph{\bibinfo{journal}{arXiv}} \bibinfo{pages}{arXiv--1911}
  (\bibinfo{year}{2019}).

\bibitem{yanase2020new}
\bibinfo{author}{Yanase, T.}, \bibinfo{author}{Nishizawa, S.},
  \bibinfo{author}{Miura, H.}, \bibinfo{author}{Takemi, T.} \&
  \bibinfo{author}{Tomita, H.}
\newblock \bibinfo{title}{New critical length for the onset of self-aggregation
  of moist convection}.
\newblock \emph{\bibinfo{journal}{Geophysical Research Letters}}
  \textbf{\bibinfo{volume}{47}}, \bibinfo{pages}{e2020GL088763}
  (\bibinfo{year}{2020}).

\bibitem{moseley2020impact}
\bibinfo{author}{Moseley, C.}, \bibinfo{author}{Pscheidt, I.},
  \bibinfo{author}{Cioni, G.} \& \bibinfo{author}{Heinze, R.}
\newblock \bibinfo{title}{Impact of resolution on large-eddy simulation of
  midlatitude summertime convection}.
\newblock \emph{\bibinfo{journal}{Atmospheric Chemistry and Physics}}
  \textbf{\bibinfo{volume}{20}}, \bibinfo{pages}{2891--2910}
  (\bibinfo{year}{2020}).

\bibitem{hirt2020cold}
\bibinfo{author}{Hirt, M.}, \bibinfo{author}{Craig, G.~C.},
  \bibinfo{author}{Sch{\"a}fer, S.~A.}, \bibinfo{author}{Savre, J.} \&
  \bibinfo{author}{Heinze, R.}
\newblock \bibinfo{title}{Cold-pool-driven convective initiation: using causal
  graph analysis to determine what convection-permitting models are missing}.
\newblock \emph{\bibinfo{journal}{Quarterly Journal of the Royal Meteorological
  Society}} \textbf{\bibinfo{volume}{146}}, \bibinfo{pages}{2205--2227}
  (\bibinfo{year}{2020}).

\bibitem{chen1997diurnal}
\bibinfo{author}{Chen, S.~S.} \& \bibinfo{author}{Houze, R.~A.}
\newblock \bibinfo{title}{Diurnal variation and life-cycle of deep convective
  systems over the tropical pacific warm pool}.
\newblock \emph{\bibinfo{journal}{Quarterly Journal of the Royal Meteorological
  Society}} \textbf{\bibinfo{volume}{123}}, \bibinfo{pages}{357--388}
  (\bibinfo{year}{1997}).

\bibitem{dai2001global}
\bibinfo{author}{Dai, A.}
\newblock \bibinfo{title}{Global precipitation and thunderstorm frequencies.
  part ii: Diurnal variations}.
\newblock \emph{\bibinfo{journal}{Journal of Climate}}
  \textbf{\bibinfo{volume}{14}}, \bibinfo{pages}{1112--1128}
  (\bibinfo{year}{2001}).

\bibitem{kawai2007diurnal}
\bibinfo{author}{Kawai, Y.} \& \bibinfo{author}{Wada, A.}
\newblock \bibinfo{title}{Diurnal sea surface temperature variation and its
  impact on the atmosphere and ocean: A review}.
\newblock \emph{\bibinfo{journal}{Journal of Oceanography}}
  \textbf{\bibinfo{volume}{63}}, \bibinfo{pages}{721--744}
  (\bibinfo{year}{2007}).

\bibitem{suzuki2009diurnal}
\bibinfo{author}{Suzuki, T.}
\newblock \bibinfo{title}{{Diurnal cycle of deep convection in super clusters
  embedded in the Madden-Julian Oscillation}}.
\newblock \emph{\bibinfo{journal}{Journal of Geophysical Research:
  Atmospheres}} \textbf{\bibinfo{volume}{114}} (\bibinfo{year}{2009}).

\bibitem{bellenger2009analysis}
\bibinfo{author}{Bellenger, H.} \& \bibinfo{author}{Duvel, J.-P.}
\newblock \bibinfo{title}{An analysis of tropical ocean diurnal warm layers}.
\newblock \emph{\bibinfo{journal}{Journal of Climate}}
  \textbf{\bibinfo{volume}{22}}, \bibinfo{pages}{3629--3646}
  (\bibinfo{year}{2009}).

\bibitem{bellenger2010role}
\bibinfo{author}{Bellenger, H.}, \bibinfo{author}{Takayabu, Y.},
  \bibinfo{author}{Ushiyama, T.} \& \bibinfo{author}{Yoneyama, K.}
\newblock \bibinfo{title}{Role of diurnal warm layers in the diurnal cycle of
  convection over the tropical indian ocean during mismo}.
\newblock \emph{\bibinfo{journal}{Monthly Weather Review}}
  \textbf{\bibinfo{volume}{138}}, \bibinfo{pages}{2426--2433}
  (\bibinfo{year}{2010}).

\bibitem{peatman2014propagation}
\bibinfo{author}{Peatman, S.~C.}, \bibinfo{author}{Matthews, A.~J.} \&
  \bibinfo{author}{Stevens, D.~P.}
\newblock \bibinfo{title}{Propagation of the madden--julian oscillation through
  the maritime continent and scale interaction with the diurnal cycle of
  precipitation}.
\newblock \emph{\bibinfo{journal}{Quarterly Journal of the Royal Meteorological
  Society}} \textbf{\bibinfo{volume}{140}}, \bibinfo{pages}{814--825}
  (\bibinfo{year}{2014}).

\bibitem{tan2015increases}
\bibinfo{author}{Tan, J.}, \bibinfo{author}{Jakob, C.},
  \bibinfo{author}{Rossow, W.~B.} \& \bibinfo{author}{Tselioudis, G.}
\newblock \bibinfo{title}{Increases in tropical rainfall driven by changes in
  frequency of organized deep convection}.
\newblock \emph{\bibinfo{journal}{Nature}} \textbf{\bibinfo{volume}{519}},
  \bibinfo{pages}{451--454} (\bibinfo{year}{2015}).

\bibitem{schumacher2020formation}
\bibinfo{author}{Schumacher, R.~S.} \& \bibinfo{author}{Rasmussen, K.~L.}
\newblock \bibinfo{title}{The formation, character and changing nature of
  mesoscale convective systems}.
\newblock \emph{\bibinfo{journal}{Nature Reviews Earth \& Environment}}
  \textbf{\bibinfo{volume}{1}}, \bibinfo{pages}{300--314}
  (\bibinfo{year}{2020}).

\bibitem{houze2004mesoscale}
\bibinfo{author}{Houze~Jr, R.~A.}
\newblock \bibinfo{title}{Mesoscale convective systems}.
\newblock \emph{\bibinfo{journal}{Reviews of Geophysics}}
  \textbf{\bibinfo{volume}{42}} (\bibinfo{year}{2004}).

\bibitem{westra2014future}
\bibinfo{author}{Westra, S.} \emph{et~al.}
\newblock \bibinfo{title}{Future changes to the intensity and frequency of
  short-duration extreme rainfall}.
\newblock \emph{\bibinfo{journal}{Reviews of Geophysics}}
  \textbf{\bibinfo{volume}{52}}, \bibinfo{pages}{522--555}
  (\bibinfo{year}{2014}).

\bibitem{prein2017simulating}
\bibinfo{author}{Prein, A.~F.} \emph{et~al.}
\newblock \bibinfo{title}{{Simulating North American mesoscale convective
  systems with a convection-permitting climate model}}.
\newblock \emph{\bibinfo{journal}{Climate Dynamics}} \bibinfo{pages}{1--16}
  (\bibinfo{year}{2017}).

\bibitem{fritsch2004improving}
\bibinfo{author}{Fritsch, J.~M.} \& \bibinfo{author}{Carbone, R.}
\newblock \bibinfo{title}{Improving quantitative precipitation forecasts in the
  warm season: A uswrp research and development strategy}.
\newblock \emph{\bibinfo{journal}{Bulletin of the American Meteorological
  Society}} \textbf{\bibinfo{volume}{85}}, \bibinfo{pages}{955--966}
  (\bibinfo{year}{2004}).

\bibitem{sukovich2014extreme}
\bibinfo{author}{Sukovich, E.~M.}, \bibinfo{author}{Ralph, F.~M.},
  \bibinfo{author}{Barthold, F.~E.}, \bibinfo{author}{Reynolds, D.~W.} \&
  \bibinfo{author}{Novak, D.~R.}
\newblock \bibinfo{title}{Extreme quantitative precipitation forecast
  performance at the weather prediction center from 2001 to 2011}.
\newblock \emph{\bibinfo{journal}{Weather and forecasting}}
  \textbf{\bibinfo{volume}{29}}, \bibinfo{pages}{894--911}
  (\bibinfo{year}{2014}).

\bibitem{liu1998numerical}
\bibinfo{author}{Liu, C.} \& \bibinfo{author}{Moncrieff, M.~W.}
\newblock \bibinfo{title}{A numerical study of the diurnal cycle of tropical
  oceanic convection}.
\newblock \emph{\bibinfo{journal}{Journal of the Atmospheric Sciences}}
  \textbf{\bibinfo{volume}{55}}, \bibinfo{pages}{2329--2344}
  (\bibinfo{year}{1998}).

\bibitem{tian2006modulation}
\bibinfo{author}{Tian, B.}, \bibinfo{author}{Waliser, D.~E.} \&
  \bibinfo{author}{Fetzer, E.~J.}
\newblock \bibinfo{title}{{Modulation of the diurnal cycle of tropical deep
  convective clouds by the MJO}}.
\newblock \emph{\bibinfo{journal}{Geophysical Research Letters}}
  \textbf{\bibinfo{volume}{33}} (\bibinfo{year}{2006}).

\bibitem{cronin2015island}
\bibinfo{author}{Cronin, T.~W.}, \bibinfo{author}{Emanuel, K.~A.} \&
  \bibinfo{author}{Molnar, P.}
\newblock \bibinfo{title}{Island precipitation enhancement and the diurnal
  cycle in radiative-convective equilibrium}.
\newblock \emph{\bibinfo{journal}{Quarterly Journal of the Royal Meteorological
  Society}} \textbf{\bibinfo{volume}{141}}, \bibinfo{pages}{1017--1034}
  (\bibinfo{year}{2015}).

\bibitem{ruppert2018diurnal}
\bibinfo{author}{Ruppert~Jr, J.~H.} \& \bibinfo{author}{Hohenegger, C.}
\newblock \bibinfo{title}{Diurnal circulation adjustment and organized deep
  convection}.
\newblock \emph{\bibinfo{journal}{Journal of Climate}}
  \textbf{\bibinfo{volume}{31}}, \bibinfo{pages}{4899--4916}
  (\bibinfo{year}{2018}).

\bibitem{ruppert2019diurnal}
\bibinfo{author}{Ruppert~Jr, J.~H.} \& \bibinfo{author}{O'Neill, M.~E.}
\newblock \bibinfo{title}{Diurnal cloud and circulation changes in simulated
  tropical cyclones}.
\newblock \emph{\bibinfo{journal}{Geophysical Research Letters}}
  \textbf{\bibinfo{volume}{46}}, \bibinfo{pages}{502--511}
  (\bibinfo{year}{2019}).

\bibitem{haerter2020diurnal}
\bibinfo{author}{Haerter, J.~O.}, \bibinfo{author}{Meyer, B.} \&
  \bibinfo{author}{Nissen, S.~B.}
\newblock \bibinfo{title}{Diurnal self-aggregation}.
\newblock \emph{\bibinfo{journal}{npj Climate and Atmospheric Science}}
  \textbf{\bibinfo{volume}{3}}, \bibinfo{pages}{30} (\bibinfo{year}{2020}).

\bibitem{lenderink2008increase}
\bibinfo{author}{Lenderink, G.} \& \bibinfo{author}{Van~Meijgaard, E.}
\newblock \bibinfo{title}{Increase in hourly precipitation extremes beyond
  expectations from temperature changes}.
\newblock \emph{\bibinfo{journal}{Nature Geoscience}}
  \textbf{\bibinfo{volume}{1}}, \bibinfo{pages}{511--514}
  (\bibinfo{year}{2008}).

\bibitem{bony2015clouds}
\bibinfo{author}{Bony, S.} \emph{et~al.}
\newblock \bibinfo{title}{Clouds, circulation and climate sensitivity}.
\newblock \emph{\bibinfo{journal}{Nature Geoscience}}
  \textbf{\bibinfo{volume}{8}}, \bibinfo{pages}{261--268}
  (\bibinfo{year}{2015}).

\bibitem{weller1996surface}
\bibinfo{author}{Weller, R.} \& \bibinfo{author}{Anderson, S.}
\newblock \bibinfo{title}{Surface meteorology and air-sea fluxes in the western
  equatorial pacific warm pool during the toga coupled ocean-atmosphere
  response experiment}.
\newblock \emph{\bibinfo{journal}{Journal of Climate}}
  \textbf{\bibinfo{volume}{9}}, \bibinfo{pages}{1959--1990}
  (\bibinfo{year}{1996}).

\bibitem{johnson1999trimodal}
\bibinfo{author}{Johnson, R.~H.}, \bibinfo{author}{Rickenbach, T.~M.},
  \bibinfo{author}{Rutledge, S.~A.}, \bibinfo{author}{Ciesielski, P.~E.} \&
  \bibinfo{author}{Schubert, W.~H.}
\newblock \bibinfo{title}{Trimodal characteristics of tropical convection}.
\newblock \emph{\bibinfo{journal}{Journal of Climate}}
  \textbf{\bibinfo{volume}{12}}, \bibinfo{pages}{2397--2418}
  (\bibinfo{year}{1999}).

\bibitem{berg2013strong}
\bibinfo{author}{Berg, P.}, \bibinfo{author}{Moseley, C.} \&
  \bibinfo{author}{Haerter, J.~O.}
\newblock \bibinfo{title}{Strong increase in convective precipitation in
  response to higher temperatures}.
\newblock \emph{\bibinfo{journal}{Nature Geoscience}}
  \textbf{\bibinfo{volume}{6}}, \bibinfo{pages}{181--185}
  (\bibinfo{year}{2013}).

\bibitem{moseley2016intensification}
\bibinfo{author}{Moseley, C.}, \bibinfo{author}{Hohenegger, C.},
  \bibinfo{author}{Berg, P.} \& \bibinfo{author}{Haerter, J.~O.}
\newblock \bibinfo{title}{Intensification of convective extremes driven by
  cloud--cloud interaction}.
\newblock \emph{\bibinfo{journal}{Nature Geoscience}}
  \textbf{\bibinfo{volume}{9}}, \bibinfo{pages}{748} (\bibinfo{year}{2016}).

\bibitem{smagorinsky1963general}
\bibinfo{author}{Smagorinsky, J.}
\newblock \bibinfo{title}{General circulation experiments with the primitive
  equations: I. the basic experiment}.
\newblock \emph{\bibinfo{journal}{Monthly Weather Review}}
  \textbf{\bibinfo{volume}{91}}, \bibinfo{pages}{99--164}
  (\bibinfo{year}{1963}).

\bibitem{pincus2009monte}
\bibinfo{author}{Pincus, R.} \& \bibinfo{author}{Stevens, B.}
\newblock \bibinfo{title}{{M}onte {C}arlo spectral integration: {A} consistent
  approximation for radiative transfer in large eddy simulations}.
\newblock \emph{\bibinfo{journal}{Journal of Advances in Modeling Earth
  Systems}} \textbf{\bibinfo{volume}{1}} (\bibinfo{year}{2009}).

\bibitem{stevens2005evaluation}
\bibinfo{author}{Stevens, B.} \emph{et~al.}
\newblock \bibinfo{title}{Evaluation of large-eddy simulations via observations
  of nocturnal marine stratocumulus}.
\newblock \emph{\bibinfo{journal}{Monthly Weather Review}}
  \textbf{\bibinfo{volume}{133}}, \bibinfo{pages}{1443--1462}
  (\bibinfo{year}{2005}).

\bibitem{seifert2006two}
\bibinfo{author}{Seifert, A.} \& \bibinfo{author}{Beheng, K.}
\newblock \bibinfo{title}{A two-moment cloud microphysics parameterization for
  mixed-phase clouds. {P}art 1: Model description}.
\newblock \emph{\bibinfo{journal}{Meteorology and {A}tmospheric {P}hysics}}
  \textbf{\bibinfo{volume}{92}}, \bibinfo{pages}{45--66}
  (\bibinfo{year}{2006}).

\bibitem{stull2012introduction}
\bibinfo{author}{Stull, R.~B.}
\newblock \emph{\bibinfo{title}{An introduction to boundary layer
  meteorology}}, vol.~\bibinfo{volume}{13} (\bibinfo{publisher}{Springer
  Science \& Business Media}, \bibinfo{year}{2012}).

\end{thebibliography}

\pagebreak
\clearpage
\newpage

\onecolumn

\renewcommand{\theequation}{S\arabic{equation}}
\renewcommand{\thesection}{S\arabic{section}}
\renewcommand{\thefigure}{S\arabic{figure}}
\renewcommand{\thetable}{S\arabic{table}}

\setcounter{equation}{0}
\setcounter{figure}{0}
\setcounter{section}{0}
\setcounter{table}{0}

\section*{\Large The diurnal path to convective self-aggregation}\label{sec:supp}
\noindent
% {\bf \large --- Supplementary Information ---}
\section*{--- Supplementary Information ---}
\noindent
In this supplementary information, we provide further analysis of interest to some expert readers, but not directly relevant for the understanding of the main text.

\vspace{1cm}

\begin{figure*}[ht]
    \begin{overpic}{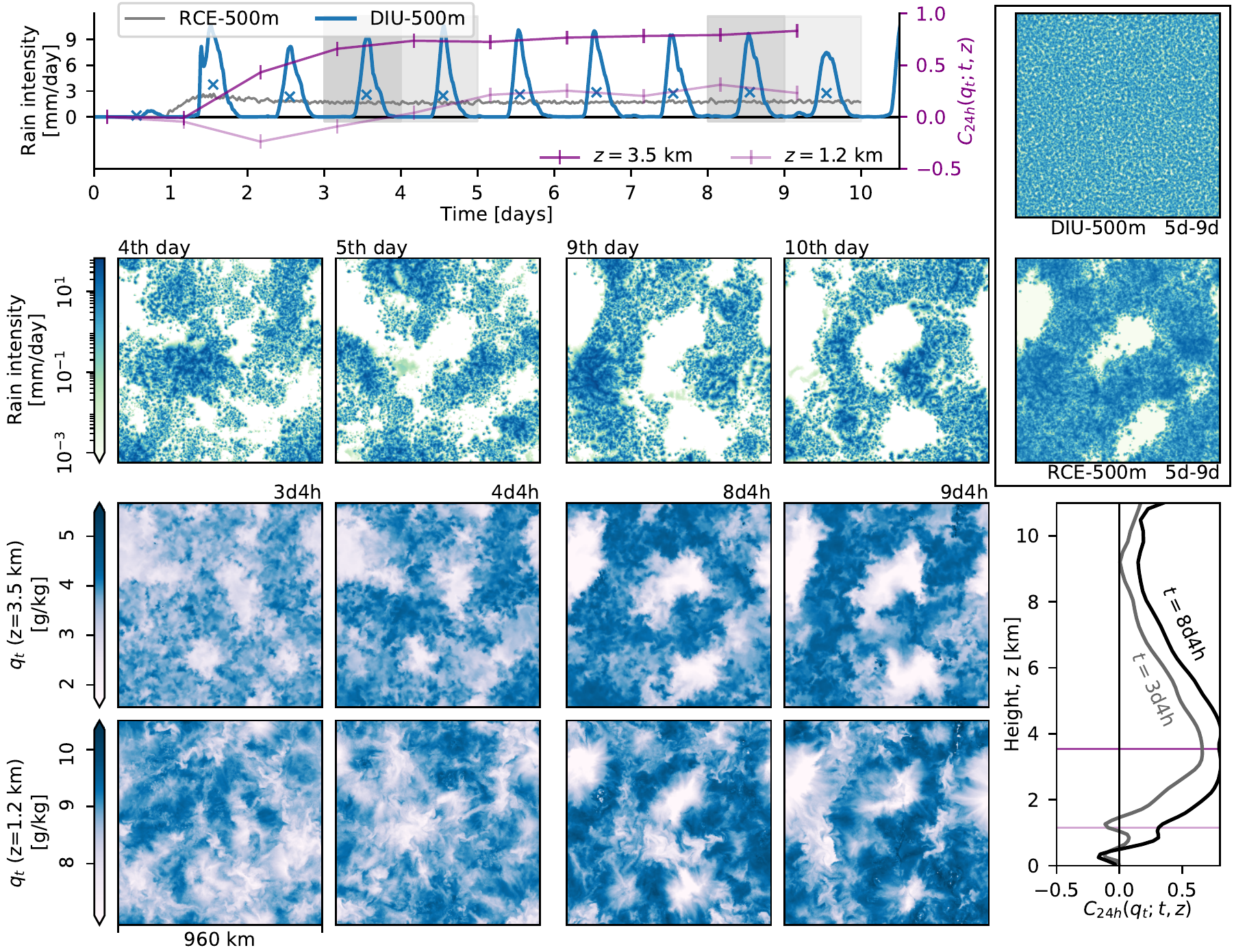}
        % Add horizontal scale arrows 
        \linethickness{1pt}	
        \put(9.5,1){\color{black}\line(1,0){5}}   % Left
        \put(9.5,1){\color{black}\line(1,0.3){1}}
        \put(9.5,1){\color{black}\line(1,-0.3){1}}
        \put(26,1){\color{black}\line(-1,0){5}}   % Right
        \put(26,1){\color{black}\line(-1,0.3){1}}
        \put(26,1){\color{black}\line(-1,-0.3){1}}
        % correct label
        \linethickness{10pt}
        \put(15,1){\color{white}\line(1,0){6}}
        \put(15,.3){480 km}
    \end{overpic}
    \caption{
    \textbf{Spatio-temporal organisation by diurnal surface temperature oscillations (DIU-500m).}
    Analogous to Fig.~\ref{fig:DIU-1km}, but for DIU-500m and RCE-500m.
    \label{fig:DIU-500m}
    }
\end{figure*}

\begin{figure}[!]
    \centering
    \begin{overpic}[trim={0 3mm 0 0},clip]{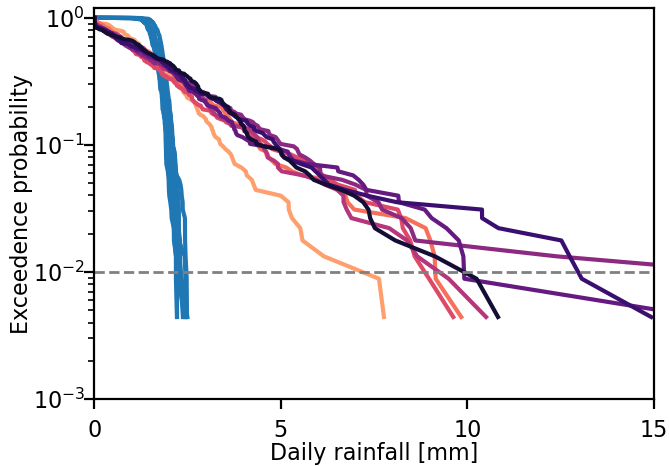}
        \put(2,62){\textbf{a}}
        \put(85,60){500m}
    \end{overpic}
    \begin{overpic}[trim={0 3mm 0 0},clip]{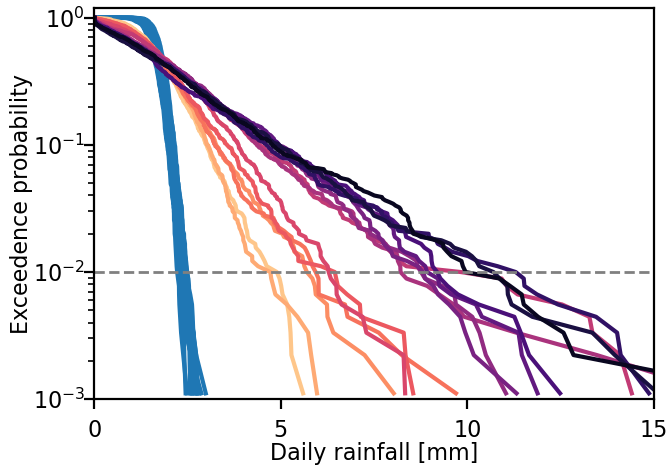}
        \put(2,62){\textbf{b}}
        \put(85,60){1km}
    \end{overpic} \\
    \vspace{3mm}
    \begin{overpic}{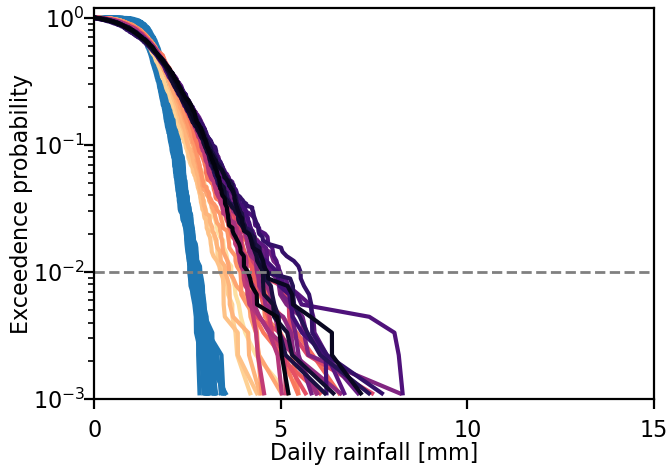}
        \put(2,65){\textbf{c}}
        \put(85,63){2km}
    \end{overpic}
    \begin{overpic}{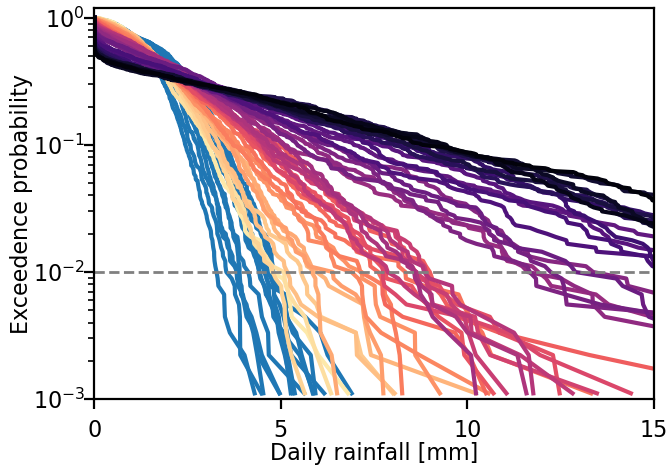}
        \put(2,65){\textbf{d}}
        \put(85,63){4km}
    \end{overpic}
    \caption{
        \textbf{Daily accumulated precipitation distribution.}
        The surface rainfall field $R$ is coarse-grained to boxes of 32~km$\times$32~km$\times$24~hours ({\it Details}: Methods), representing daily rainfall averages with squares of 32~km side length. 
        For various simulation day, we plot the normalised probability of  such daily rainfall averages to lie above any given threshold.
        Curves of colours in the sequence of yellow-red-purple-black represent distribution functions for consecutive days.
        %for each of the individual simulation days.
        Analogously, the blue curves represent the RCE experiments.
        %The yellow-orange-pink-purple-black colours corresponds to days in the DIU experiments, incriminating from light to dark.
        The first 48 hours are omitted in all datasets, to discard the spin-up period.
        Panels correspond to horizontal resolutions of: \textbf{a}, 500m; \textbf{b}, 1~km; \textbf{c}, 2~km; \textbf{d}, 4~km.
        % \textbf{a} RCE-simlation 960x960 pixels, 1km horizontal resolution, surface temperature 298K (RCE).
        % \textbf{b} Similar, but with a 5K diurnal surface temperature oscillation amplitude (DIU).
        In all panels, the grey, dashed horizontal lines mark the 99'th percentiles, as a typical measure of extreme precipitation.
    Note the pronounced increase in extreme precipitation for the later days of the DIU-simulation, where the 99'th percentile reaches a multiple of that for the RCE-simulation.
    For all panels, note the logarithmic vertical axis scaling.
    %\textcolor{red}{Curves represent all complete days, except the first 48 hours in each experiment.}
    }
    \label{fig:precipitation}
\end{figure}

\begin{figure}
    \centering
    \includegraphics[width=1\textwidth]{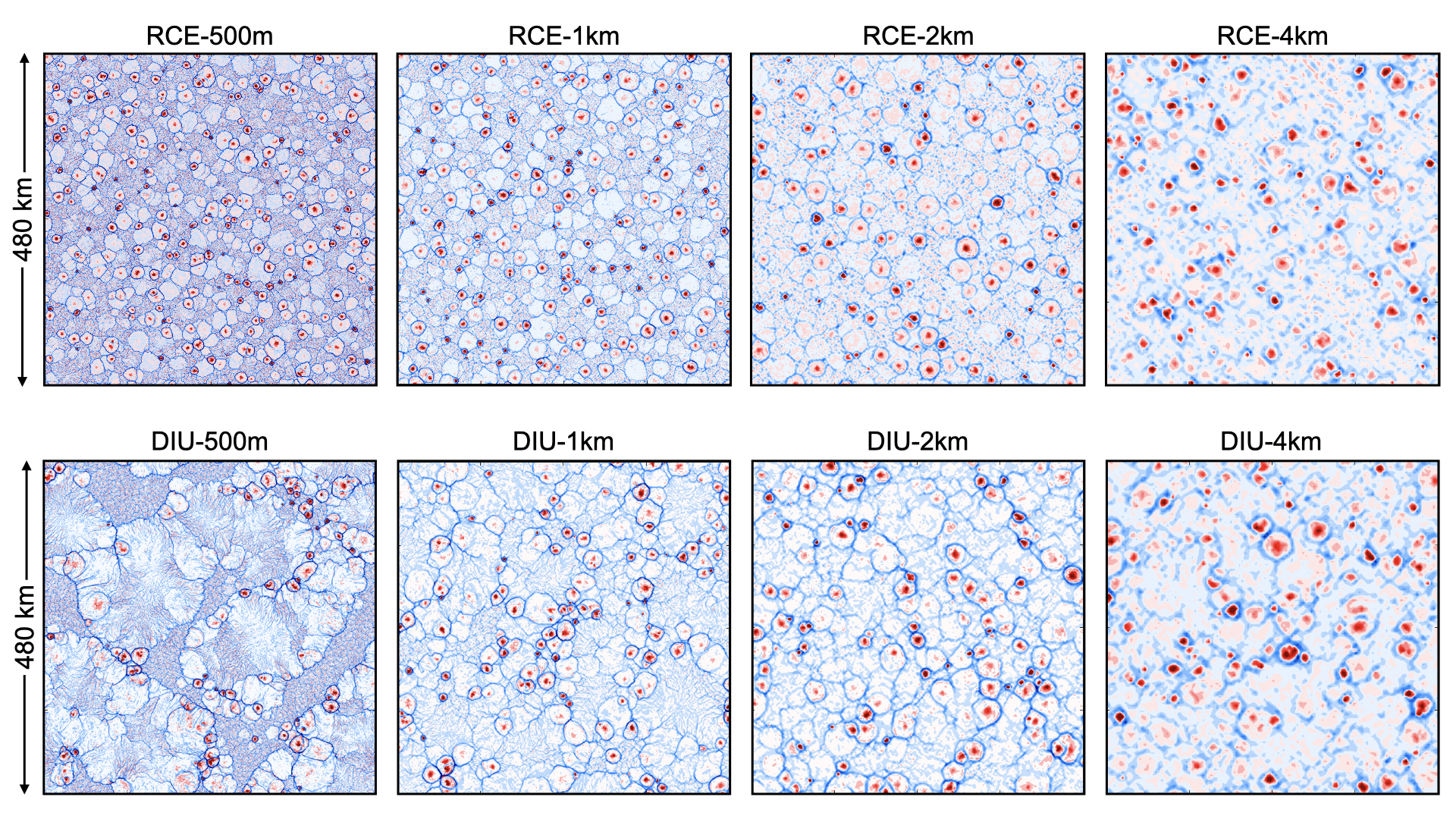}
    \caption{\textbf{Contours of surface horizontal velocity divergence fields}. Snapshots of $\frac{\partial u}{\partial y} + \frac{\partial v}{\partial x}$ are extracted from the first level (z=50m) at 18h00 on the sixth day. The contours range from -0.004 to 0.004 s$^{-1}$ (blue to red) and are presented from left to right in decreasing order of spatial resolution for both (top) RCE and (bottom) diurnal configurations.}
    \label{fig:2Ddivergence_allCases}
\end{figure}

\begin{figure}[!]
%\vspace{6cm}
    \centering
    \begin{overpic}[width=0.6\textwidth]{diu2rce/dummy.pdf}
    %\put(0,173){\textbf{a}}
    \put(0,26){\includegraphics[trim=0 2.3cm 0 0,clip,width=.55\textwidth]{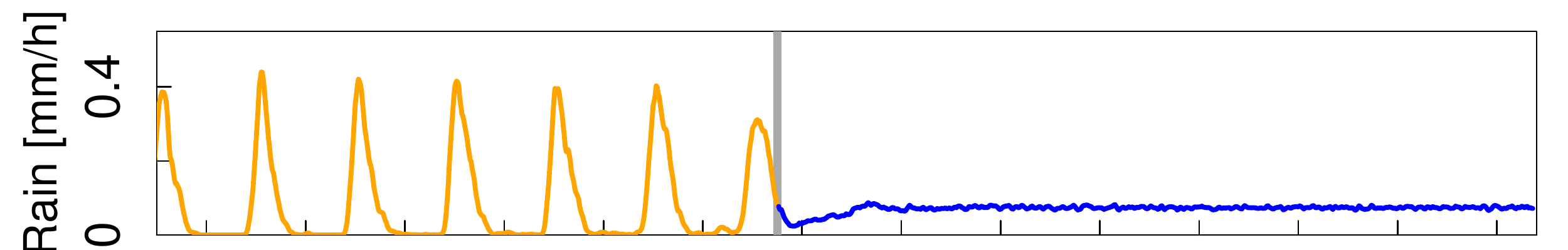}}
    \put(0,16){\includegraphics[trim=0 2.3cm 0 0,clip,width=.55\textwidth]{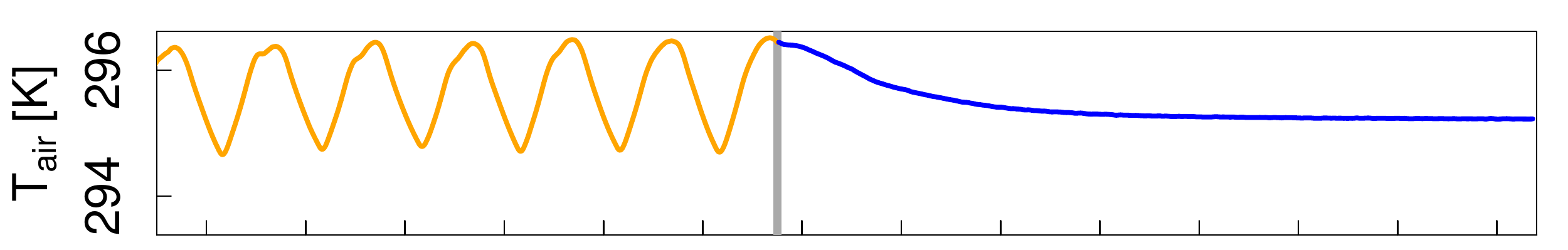}}
     \put(0,0){\includegraphics[trim=0 0cm 0 0,clip,width=.55\textwidth]{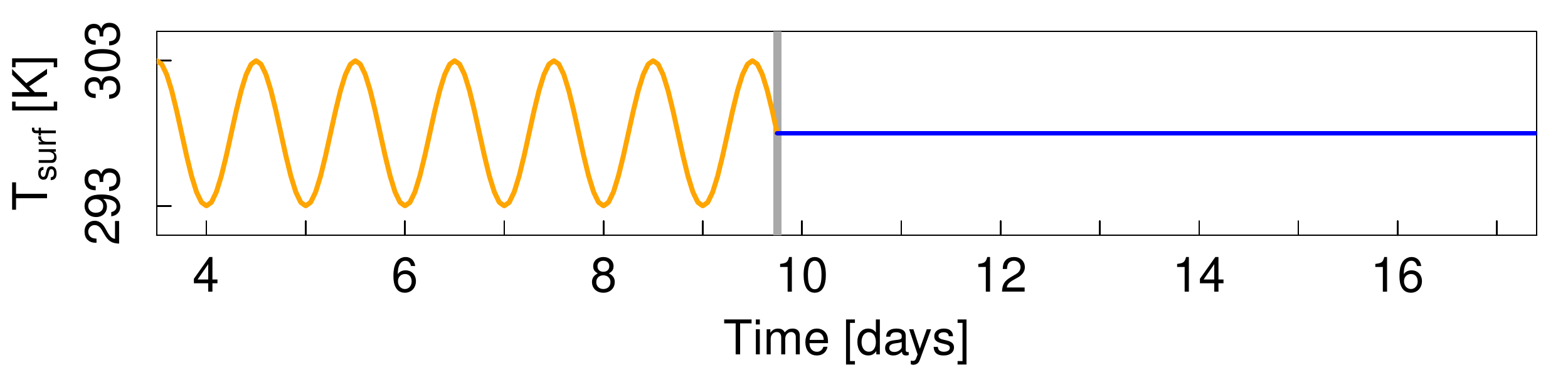}}
     \linethickness{1.4pt}	
    \put(18,35){\color{black}\line(1,0){30}}
    \put(48,35){\color{black}\line(-1,0.4){2}}
    \put(48,35){\color{black}\line(-1,-0.4){2}}
    \put(20,36){DIU}
    \put(36,36){RCE}    
    
\end{overpic}
\caption{{\bf Domain mean time series for DIU2RCE-500m.}
Time series of horizontally-averaged rainfall intensity, near-surface air temperature ($T(z=50\,m)$) and prescribed surface temperature ($T_{surf}$).
Note the transition between temporally-varying (DIU) and temporally constant (RCE) surface temperature at $t=$9d18h, as well as the response in $T_{air}$ and rainfall.
}
\label{fig:domain_mean_timeseries}
\end{figure}

%\vspace{1cm}

\begin{figure}[ht]
\vspace{2cm}
    \centering
    \begin{overpic}[width=0.4\textwidth]{diu2rce/dummy.pdf}
%    \put(-38,173){\textbf{a}}
%    \put(-45,159){\includegraphics[trim=0 2.3cm 0 0,clip,width=.5\textwidth]{diu2rce/timeseries_r_int_diu2rce_fldmean.pdf}}
%    \put(-45,146){\includegraphics[trim=0 2.3cm 0 0,clip,width=.5\textwidth]{diu2rce/timeseries_t50m_diu2rce_fldmean.pdf}}
%     \put(-45,125){\includegraphics[trim=0 0cm 0 0,clip,width=.5\textwidth]{diu2rce/timeseries_tsurf_diu2rce_fldmean.pdf}}

    \put(-53,84.5){\includegraphics[trim=1cm 0 0 0, clip, width=.08\textwidth]{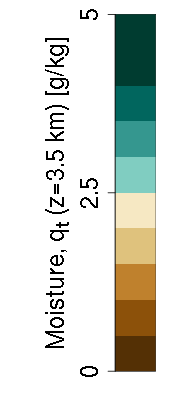}}
    
    \put(-40,86){\includegraphics[width=.22\textwidth]{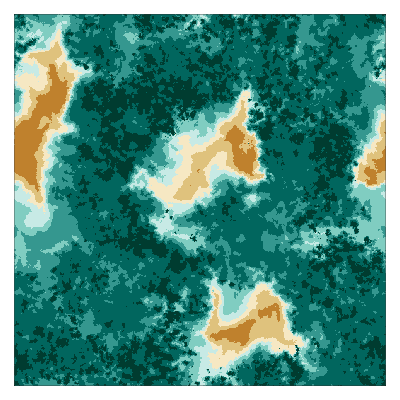}}
        \put(-12,124){9d18h}
    \put(0,86){\includegraphics[width=.22\textwidth]{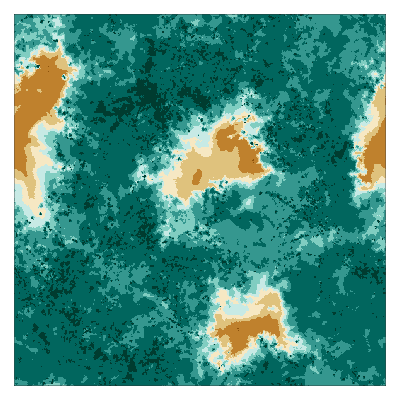}}
    \put(2,124){\textbf{b}}
    \put(26,124){10d18h}
        \put(40,86){\includegraphics[width=.22\textwidth]{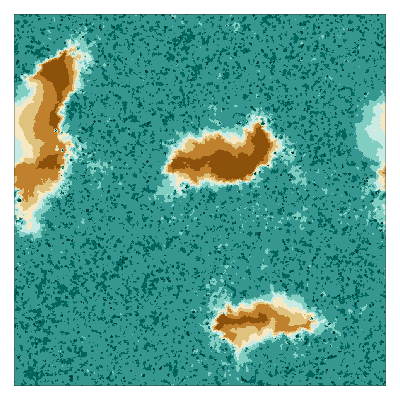}}
    \put(42,124){\textbf{c}}
    \put(68,124){17d8h}
    \put(80,83.5){\includegraphics[width=.22\textwidth]{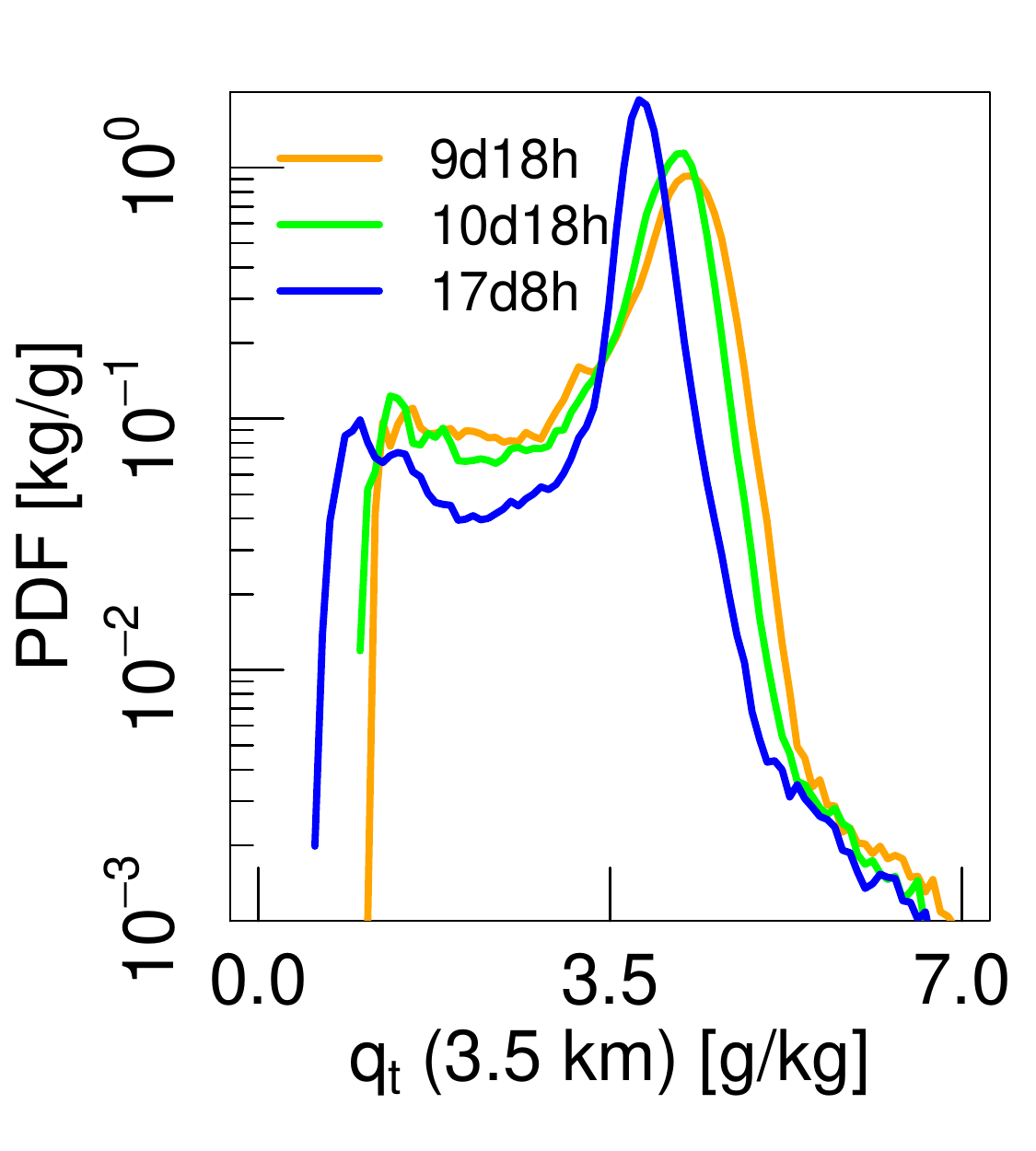}}
    \put(88,124){\textbf{d}}
    
    \put(-53,44.5){\includegraphics[trim=1cm 0 0 0, clip, width=.08\textwidth]{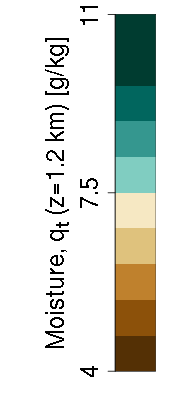}}
    \put(-40,46){\includegraphics[width=.22\textwidth]{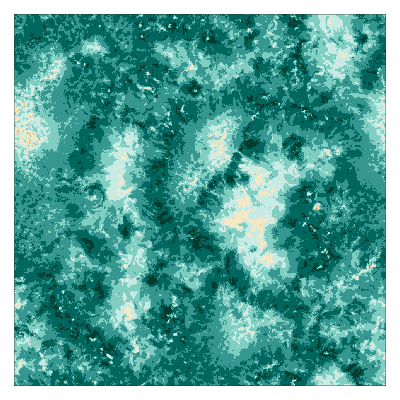}}
    \put(0,46){\includegraphics[width=.22\textwidth]{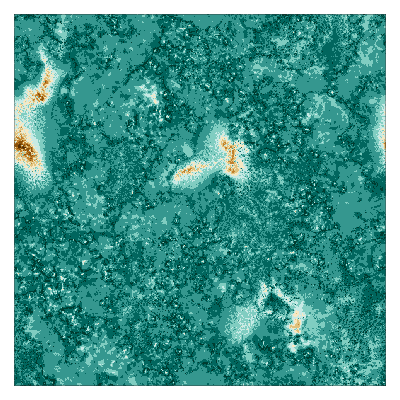}}
        \put(40,46){\includegraphics[width=.22\textwidth]{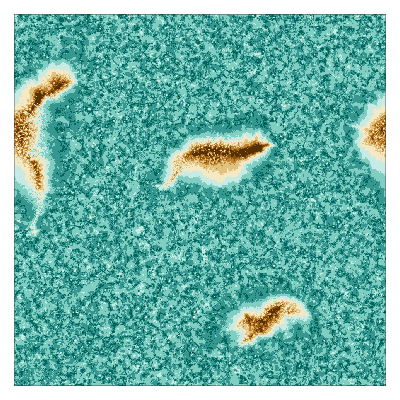}}
    \put(-38,84){\textbf{e}}
    \put(80,43.5){\includegraphics[width=.22\textwidth]{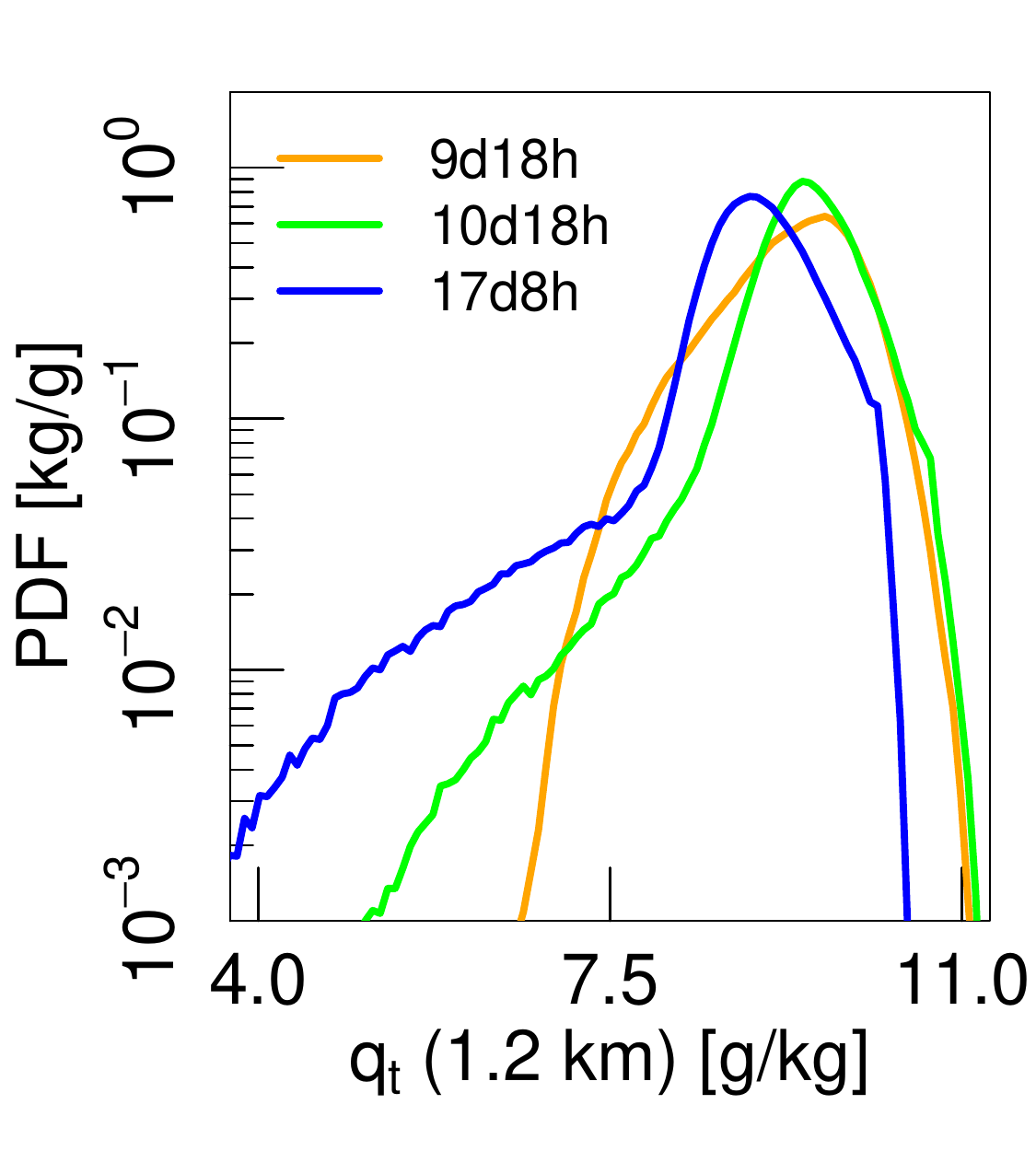}}
    %\linethickness{1pt}	
    %\put(-38.5,46){\color{black}\line(1,0){10}}
    %\put(-13,46){\color{black}\line(1,0){10}}
    %\put(-38.5,46){\color{black}\line(1,0.3){1}}
    %\put(-38.5,46){\color{black}\line(1,-0.3){1}}
    %\put(-3,46){\color{black}\line(-1,0.3){1}}
    %\put(-3,46){\color{black}\line(-1,-0.3){1}}
    %\put(-26,44.5){480 km}
    
    \linethickness{3pt}	
    \put(-.8,-37.5){\color{gray}\line(0,1){162}}
     %   \linethickness{1.4pt}	
    %\put(-15,71){\color{black}\line(1,0){30}}
    %\put(15,71){\color{black}\line(-1,0.4){2}}
    %\put(15,71){\color{black}\line(-1,-0.4){2}}
    %\put(-22,70){DIU}
    %\put(17,70){RCE}    
%
%
%
   \put(-53,1){\includegraphics[trim=1cm 0 0 0, clip, width=.08\textwidth]{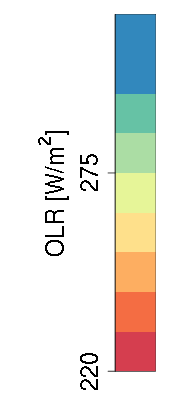}}
   %\put(-53,44.5){\includegraphics[trim=1cm 0 0 0, clip, width=.08\textwidth]{diu2rce/colorbar_horiplot_anomaly_out.vol.q.LEV30_diurnal_500m_diu2rce_t1=1_t2=1.png}}
    \put(-40,2.5){\includegraphics[width=.22\textwidth]{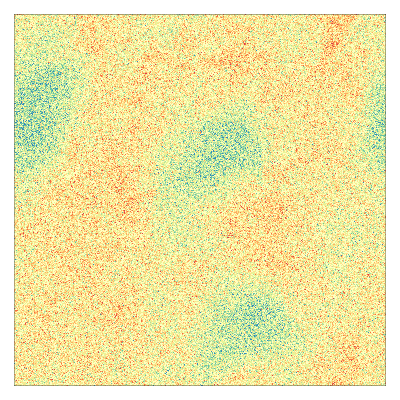}}
        \put(-38,124){\textbf{a}}
        \put(-12,41){8d18h}
    \put(0,2.5){\includegraphics[width=.22\textwidth]{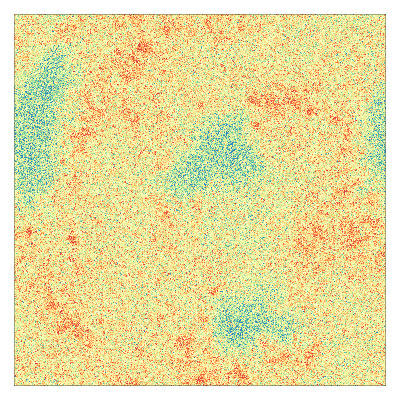}}
    \put(2,84){\textbf{f}}
    \put(26,41){10d18h}
        \put(40,2.5){\includegraphics[width=.22\textwidth]{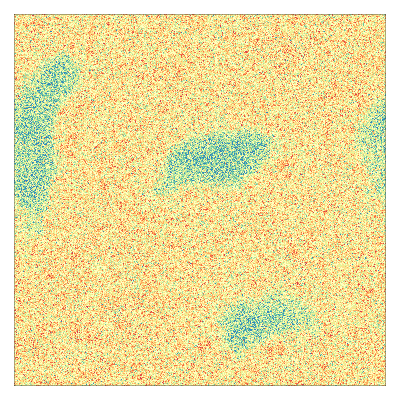}}
    
    \put(-38,41){\textbf{i}}
    \put(2,41){\textbf{j}}
    \put(42,41){\textbf{k}}
    \put(88,41){\textbf{l}}

    \put(42,84){\textbf{g}}
    \put(68,41){16d9h}
    \put(80,0){\includegraphics[width=.22\textwidth]{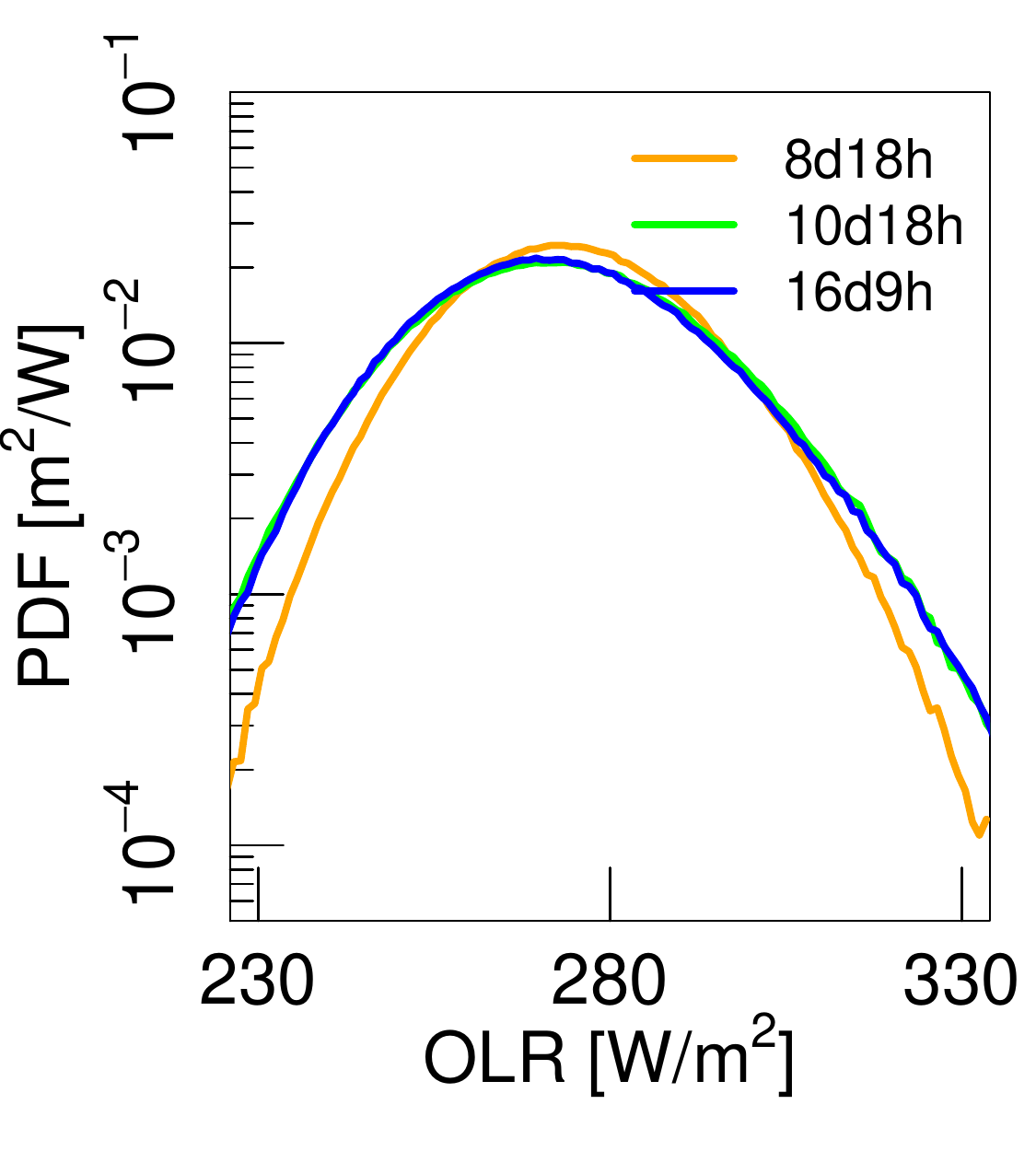}}
    \put(88,84){\textbf{h}}
    
    \put(-53,-39){\includegraphics[trim=1cm 0 0 0, clip, width=.08\textwidth]{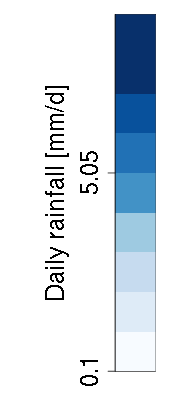}}
    \put(-40,-37.5){\includegraphics[width=.22\textwidth]{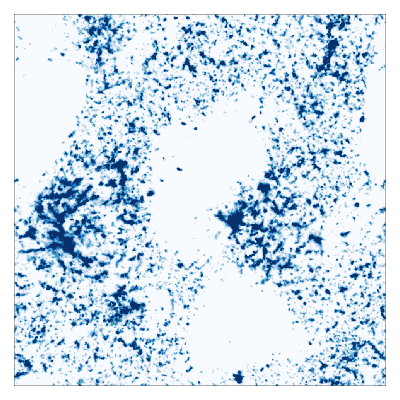}}
    \put(0,-37.5){\includegraphics[width=.22\textwidth]{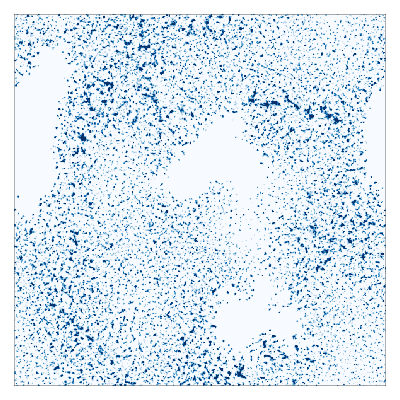}}
        \put(40,-37.5){\includegraphics[width=.22\textwidth]{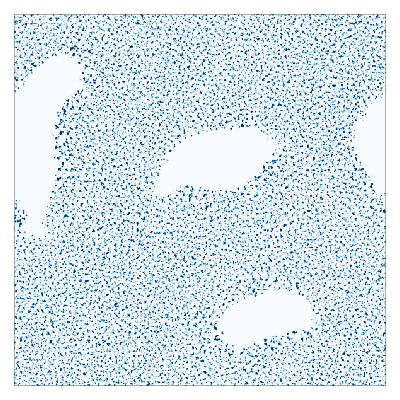}}
    
    \put(80,-40){\includegraphics[width=.22\textwidth]{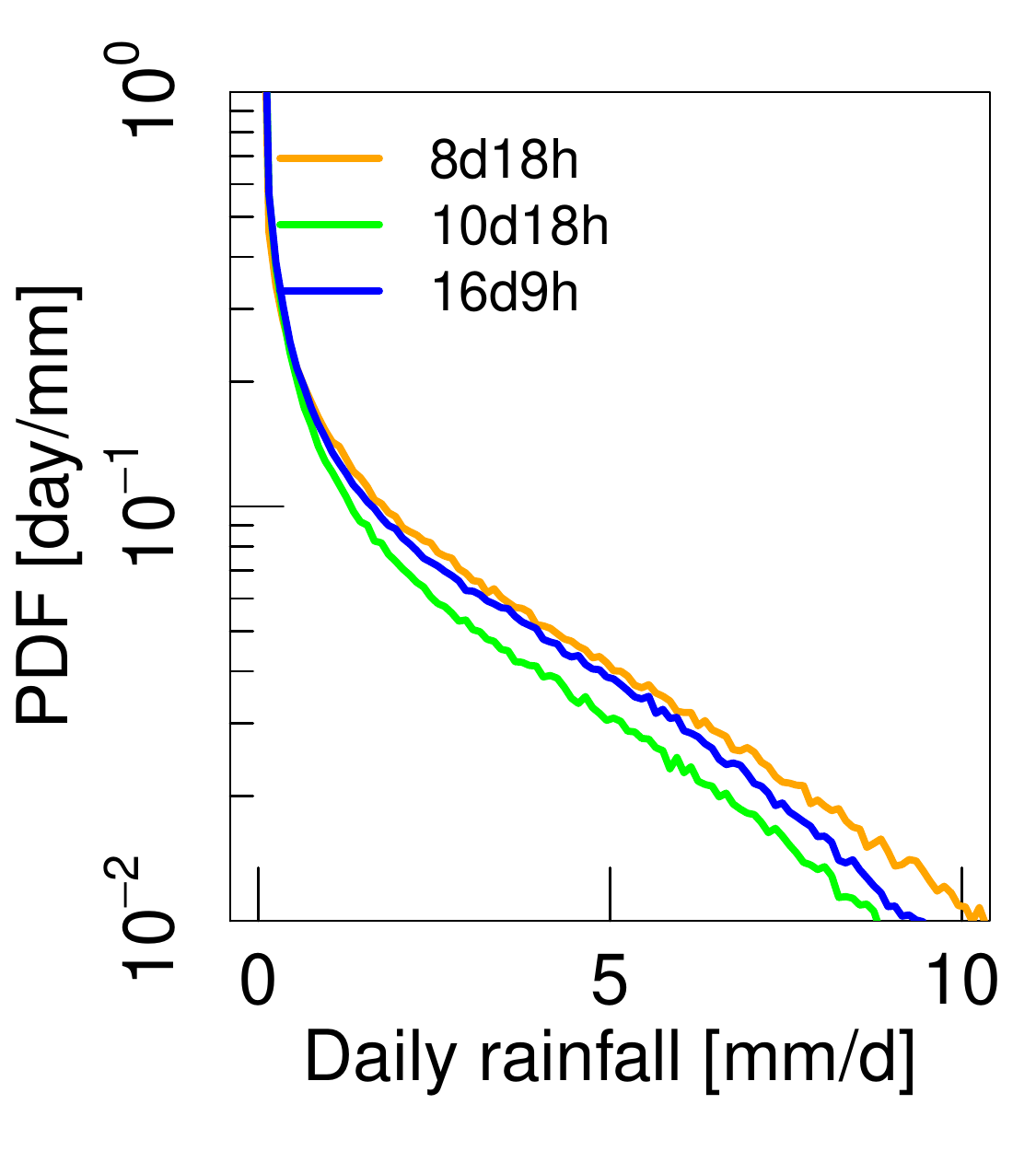}}
    \put(-38,1){\textbf{m}}
    \put(2,1){\textbf{n}}
    \put(42,1){\textbf{o}}
    \put(88,1){\textbf{p}}  
    \linethickness{1pt}	
    \put(-38.5,-38){\color{black}\line(1,0){10}}
    \put(-13,-38){\color{black}\line(1,0){10}}
    \put(-38.5,-38){\color{black}\line(1,0.3){1}}
    \put(-38.5,-38){\color{black}\line(1,-0.3){1}}
    \put(-3,-38){\color{black}\line(-1,0.3){1}}
    \put(-3,-38){\color{black}\line(-1,-0.3){1}}
    \put(-26,-39.5){480 km}
    
    % adding one more arrow on top of all
     \linethickness{1.4pt}	
    \put(-15,128){\color{black}\line(1,0){30}}
    \put(15,128){\color{black}\line(-1,0.4){2}}
    \put(15,128){\color{black}\line(-1,-0.4){2}}
    \put(-22,127){DIU}
    \put(17,127){RCE}    
    \end{overpic}
    \vspace{4.2cm}
    \caption{{\bf Transition from DIU to RCE.} 
    {\bf a}, Water vapour mixing ratio anomaly $q'(z=3.5$~km$)\equiv q(z=3.5$~km$)-\overline{q_t}(z=3.5$~km$)$, where the overline denotes the horizontal mean, at the time of transition between DIU and RCE ($t=9.75$~days) for DIU2RCE-500m. 
    %$q'(z=3.5\,km)$ was averaged temporally over the two-day period $[7.75,9.75]$ days of DIU500m. 
    {\bf b}, Analogous to (a), but one day after the transition to RCE ($t=10.75$ days).
    {\bf c}, Analogous to (a), but more than six days after the transition ($t=17.35$ days).
    {\bf d}, PDF of $q_t(z=3.5$~km$)$ for all three times shown in (a)---(c).
    Note the overall broadening and increasing bimodality of the humidity distribution function with time.
    {\bf e}---{\bf h}, analogous to (a)---(d), but for $q(z=1.2$~km$)$.
    %Note the extreme drying seen in the lower part of the distribution function in (h).
    %Take note also of the logarithmic vertical axis scaling in (d) and (h). 
    {\bf i}---{\bf k}, Two-day temporal average (centred at the times noted on the panels) of outgoing long-wave radiation (OLR) at the model top.
    {\bf d}, PDF of OLR, corresponding to the panels (i)---(k). 
    Note the widening of the distribution function.
    {\bf m}---{\bf p}, Analogous to (i)---(l) but for rainfall intensity.
    Note the pronounced spatial structure in the rain field in (m) but the gradual relaxation to a featureless rainy sub-region and a rain-free dry region (n,o).
    In all histograms the vertical axis is logarithmic.
    }
    \label{fig:diu2rce_SI}
\end{figure}

\end{document}